\documentclass[pre,amsmath]{revtex4}
\usepackage{graphicx}
\usepackage{color}
\usepackage{bm}
\usepackage{epsfig}
\usepackage{latexsym}
\usepackage{pifont}
\usepackage{float}
\usepackage[utf8]{inputenc}
\usepackage{textgreek}
\graphicspath{ {figure/} }

\begin{document}

\newcommand{\be}{\begin{equation}}
\newcommand{\ee}{\end{equation}}
\newcommand{\bea}{\begin{eqnarray}}
\newcommand{\eea}{\end{eqnarray}}
\newcommand{\nn}{\nonumber}
\title{A minimal physical model for curvotaxis driven by curved protein complexes at the cell's leading edge}
%\title{Physical model for curvotaxis driven by curved proteins at the cell's leading edge}
%\title{Cells distinguish between concave and convex curvatures while migrating on curved surfaces}

\author{$^1$Raj Kumar Sadhu\footnote{raj-kumar.sadhu@curie.fr} \footnote{Present address: Institut Curie, PSL Research University, CNRS, UMR 168, Paris, France}}
\author{$^2$Marine Luciano}
\author{$^3$Wang Xi}
\author{$^{4}$Cristina Martinez-Torres}
\author{$^5$Marcel Schr{\"{o}}der}
\author{$^5$Christoph Blum}
\author{$^5$Marco Tarantola}
%\author{$^6$Vahid Nasiri}
\author{$^6$Samo Peni\v{c}}
\author{$^6$Ale\v{s} Igli\v{c}}
\author{$^{4}$Carsten Beta}
\author{$^7$Oliver Steinbock}
\author{$^5$Eberhard Bodenschatz}
\author{$^3$Beno{\^{i}}t Ladoux}
\author{$^8$ Sylvain Gabriele}
\author{$^1$Nir S. Gov\footnote{nir.gov@weizmann.ac.il}}
\affiliation{$^1$Department of Chemical and Biological Physics, Weizmann Institute of Science, Rehovot 7610001, Israel.}
\affiliation{$^2$Department of Biochemistry, University of Geneva, Geneva Switzerland.}
\affiliation{$^3$ Universite Paris Cite, CNRS, Institut Jacques Monod, F-75013 Paris, France.}
\affiliation{$^{4}$Institute of Physics and Astronomy, University of Potsdam, Potsdam 14476, Germany.}
\affiliation{$^5$Max Planck Institute for Dynamics and Self-Organization, G{\"{o}}ttingen, Germany.}
\affiliation{$^6$Laboratory of Physics, Faculty of Electrical Engineering, University of Ljubljana, Ljubljana, Slovenia.}
\affiliation{$^7$Florida State University, Department of Chemistry and Biochemistry, Tallahassee, FL 32306-4390, USA}
\affiliation{$^8$Mechanobiology \& Biomaterials group, Interfaces and Complex Fluids Laboratory, Research Institute for Biosciences, CIRMAP, University of Mons, Mons Belgium.}

\begin{abstract}
Cells often migrate on curved surfaces inside the body, such as curved tissues, blood vessels or highly curved protrusions of other cells. Recent \textit{in-vitro} experiments provide clear evidence that motile cells are affected by the curvature of the substrate on which they migrate, preferring certain curvatures to others, termed ``curvotaxis". The origin and underlying mechanism that gives rise to this curvature sensitivity are not well understood. Here, we employ a ``minimal cell" model which is composed of a vesicle that contains curved membrane protein complexes, that exert protrusive forces on the membrane (representing the pressure due to actin polymerization). This minimal-cell model gives rise to spontaneous emergence of a motile phenotype, driven by a lamellipodia-like leading edge. By systematically screening the behaviour of this model on different types of curved substrates (sinusoidal, cylinder and tube), we show that minimal ingredients and energy terms capture the experimental data. The model recovers the observed migration on the sinusoidal substrate, where cells move along the grooves (minima), while avoiding motion along the ridges. In addition, the model predicts the tendency of cells to migrate circumferentially on convex substrates and axially on concave ones. Both of these predictions are verified experimentally, on several cell types. Altogether, our results identify the minimization of membrane-substrate adhesion energy and binding energy between the membrane protein complexes as key players of curvotaxis in cell migration. 
\end{abstract}

\maketitle

\section{Introduction}
Cell migration is an important biological process that plays a central role in immune response, wound healing, tissue homeostasis etc \cite{Howard_book,Friedl2009}. While the environment of a cell \textit{in vivo} is geometrically complex, most of the studies focus on cell spreading and migration on flat substrates \cite{Sheetz2004PRL,cavalcanti2007cell,cuvelier2007universal}. Previous studies on 2D patterned flat surfaces have shown that cells adapt their shape and their internal cytoskeleton to these 2D geometries \cite{thery2006anisotropy,mohammed2019substrate,chen2019large}. However, eukaryote cells, adhering and migrating on a solid substrate, are observed to also interact with the topography of the substrate and modify their motility \cite{song2015sinusoidal,pieuchot2018curvotaxis}. The alignment and the direction of migration of isolated cells, in response to the topography, crucially depends upon the cell type. For example, fibroblasts are found to align axially on the surface of a cylinder, while epithelial cell align circumferentially \cite{blum2015curvotaxis,Sebastien2020,Richard2019}. In another experiment \cite{song2015sinusoidal}, the migration of T-lymphocytes was studied on a surface with sinusoidal (wavy) height undulations. The cells were found to move axially in the grooves (minima) of the surface topography, avoiding migration on the ridges (maxima). In \cite{pieuchot2018curvotaxis}, the dynamics of several cell types was studied on a 2D sinusoidal surface. Adherent fibroblast cells, dominated by stress-fibers and weakly motile, were found to settle in the concave grooves or adhere aligned to the undulation axis (both on grooves and ridges)  \cite{Werner2018,Werner2019}. In many adherent cells, the alignment is found to be determined by the competition between the bending energy of the stress fibers, of the nucleus and the contractile forces \cite{Biton2009,SANZHERRERA2009,werner2020cellular}.  %The leading edge protrusions of metastatic cancer cells were observed to coil and rotate around the fiber's axis in a curvature-dependent manner \cite{mukherjee2019cancer,nain2022}. Similar coiling of protrusion is also observed in \cite{Nils2015} where `fin'-like membrane protrusions were found to coil around the fibers for several cell types (fibroblasts, epithelial, endothelial).
At the level of cell collectives, both alignment and cell migration within the confluent tissue, is found to be affected by the substrate curvature, experimentally \cite{yevick2015architecture,xi2017emergent,mazalan2020effect,luciano2021cell,Carles2022,tang2022collective} and in theoretical analysis \cite{lin2020collective}. 

Despite these studies, the underlying mechanisms that determine the response of migrating cells to the substrate curvature are still not well understood, even at the level of single migrating cells. A few theoretical studies addressed the curvature response of an isolated motile cell. One model contains a detailed description of the cellular mechanics, and is based on the assumption of a central role for the nuclear dynamics in controlling the cell migration on the curved surface \cite{vassaux2019biophysical}. A similar approach of modelling cell migration as arising from coupling the nucleus to random peripheral protrusions \cite{he2017substrate}, produced migration patterns that were in qualitative agreement with observations \cite{song2015sinusoidal}, i.e. resulting in cells migrating preferentially along the grooves. Another model provides a simpler and more general description in terms of an active-fluid \cite{winkler2019confinement}, but its predictions were not systematically compared to experiments. A similar model was proposed to describe amoeba cells moving along ridges, guided by a reaction-diffusion mechanism adapted from macropinocytic cup formation \cite{honda2021microtopographical}. 

\begin{figure}[h!]
\centering
\includegraphics[scale=1.25]{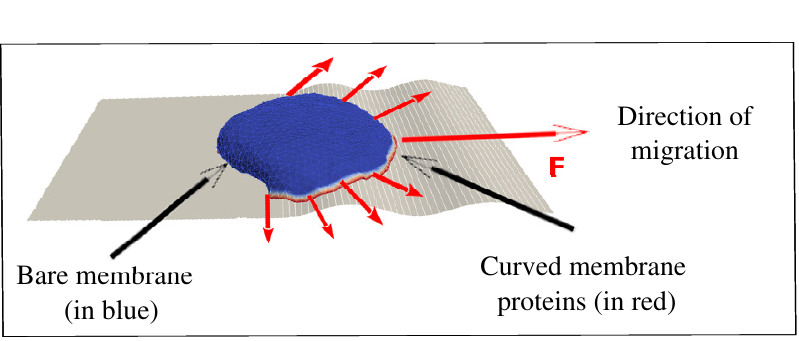}
\caption{Motile vesicle migrating on sinusoidal substrate. The red dots on the rim of the migrating vesicle are the curved membrane protein complexes with positive intrinsic curvature (convex), while the blue part represents bare membrane. The red arrows with filled arrowheads indicate the forces exerted by the curved protein complexes on the membrane, directed towards the local outwards normal. The total force is indicated by the red arrow with the empty arrowhead, which gives the net propulsion force, and direction of migration of the vesicle.}
\label{model-fig}
\end{figure}

For cell migration that is driven by the lamellipodia protrusion, understanding the migration on curved substrates requires an understanding of the mechanisms that drive the formation of the lamellipodia. Recently we have proposed a theoretical model where the lamellipodia forms as a self-organization of curved actin nucleators, coupled with adhesion to the substrate \cite{miha2019,Sadhu2021}. We showed that due to the spontaneous curvature of the actin nucleators, they aggregate at the cell-substrate contact line, and induce an outwards normal force, which represents the protrusive force due to actin polymerization. Curvature sensitive membrane complexes that contain actin nucleation factors \cite{RAMAKRISHNAN2013,schmeiser2015flatness,Rangamani2018,Luka2021} have been found in experimental observations at the leading edge of cellular protrusions \cite{Rangamani2017,joern2014,begemann2019mechanochemical,pipathsouk2021wave}. This model can give rise to the spontaneous formation of a lamellipodia-like protrusion, with a stable and asymmetric leading edge, that drives the migration of the simulated membrane vesicle (Fig.\ref{model-fig}). We found these motile vesicles to be highly persistent on a flat surface, maintaining robustly their direction of migration \cite{Sadhu2021}. 

Here, we use the spontaneously migrating vesicle that arises in our model (Fig.\ref{model-fig}), as a minimal model of a migrating cell, to explore its behavior and motility on a wide range of curved surfaces. Indeed, we explore surfaces with smooth sinusoidal shape undulations, as well as fibers (outside of cylindrical surfaces) and tubes (inside of cylinders). We do not explore here topographies with sharp edges and barriers or on length-scales much smaller than the cellular length-scale (such as these experimental studies (\cite{lee2020quantifying,brunetti2021wasp,bull2022actin}), as these will require a much finer mesh for the vesicle surface triangulation and are consequently computationally costly. In addition, sharp edges will increase the chance of our motile vesicle loosing its polarity \cite{Sadhu2021}. 

Despite the simplicity of our model, the migration patterns of our motile vesicle on the curved surfaces correspond closely to published, as well as new experimental observations that we present here, of cell migration over curved surfaces. The model vesicle is found to move perpendicular to a sinusoidal topography of short length-scale, while it tends to migrate circumferentially around fibers (and pillars). These calculated migration patterns of the motile vesicle, are used to predict the migration patterns of motile cells, and we verify these predictions in experiments using several cell types. Our minimal model for cell migration suggests that some aspects of curvotaxis, of cells that migrate using lamellipodia protrusions, can be universally explained using physical principles. 

%%%%%%%%%%%%%%%%%%%%%%%%%%%%%%%%%%%%%%%%%%%%%%%%%%%%%%%%%%%%%%%%%%%%%%%%%%%%%%%%%%%%%%%%%%%%%%%%%%%%%%%
\section{Results}
When simulating the migration of our motile vesicle (Fig.\ref{model-fig}) on curved surfaces, we have to note that our motile vesicle can easily loose its polarization and motility if it encounters large amplitude and sharp height undulations or barriers \cite{Sadhu2021}. This ``fragility" of the motile phenotype in our model constrains us to explore surfaces with small gradients of height undulations, such that our vesicle does not loose its polarization and motility. Our vesicle loses its motility when its leading edge protein cluster (Fig.\ref{model-fig}) breaks up into two or more parts, which happens when the vesicle collides with an obstacle of large height gradient, or can occur spontaneously due to noise \cite{Sadhu2021}. In our model this event is irreversible, while in real cells there are internal mechanisms that allow cells to recover their polarized shape and resume motility \cite{stites1998phosphorylation,andrew2007chemotaxis,neilson2011chemotaxis}.

We therefore explore below the migration of our motile minimal-cell system on smooth surfaces where the curvature changes gradually on the length-scale of the vesicle surface triangulation.
The shape of the sinusoidal substrate in the simulation is of the form: $z=z_m \sin (2 \pi y/y_m)$, such that the sinusoidal variations are along the y-direction and the curvature remains constant along x-direction. We use several combinations of $z_m$ and $y_m$ in our simulations: (1) $z_m=10~l_{min};~y_m=120~l_{min}$, (2) $z_m=2~l_{min};~y_m=30~l_{min}$, and (3) $z_m=1~l_{min};~y_m=15~l_{min}$. By keeping the ratio $z_m/y_m\ll1$, we remain in the regime of small undulations, which maintains the motility of our simulated vesicle. Similarly for the migration on fibers and inside tubes, we keep their radius large compare with the triangulation length-scale.
See Sec. S1 for more details regarding the theoretical model \cite{RAMAKRISHNAN2014,Sadhu2021,miha2019,sadhu_coiling,Sadhu_phagocytosis,GOV-chapter25, DRAB-chapter26}.

In the simulations with sinusoidal surface, we start with a motile vesicle, that we formed on a flat substrate, and then deform the substrate into a curved shape and allow the vesicle to evolve so that it matches the curved substrate to which it is adhered (see SI Sec. S3, Fig.S3 and Movie-{\textcolor{blue}{S1}} for more details). This allows us to control the initial direction of motion of the vesicle. Alternatively, we can start from a spherical-like vesicle and let it spread over the curved surface. However, in this case, we do not have any control over the initial direction of migration.

\subsection{Cellular migration on sinusoidal surfaces: large wavelength}

%\subsection{Cellular migration on sinusoidal substrate with $y_m/R_{vesicle} > 2$}
\begin{figure}[h!]
\centering
\includegraphics[scale=1.1]{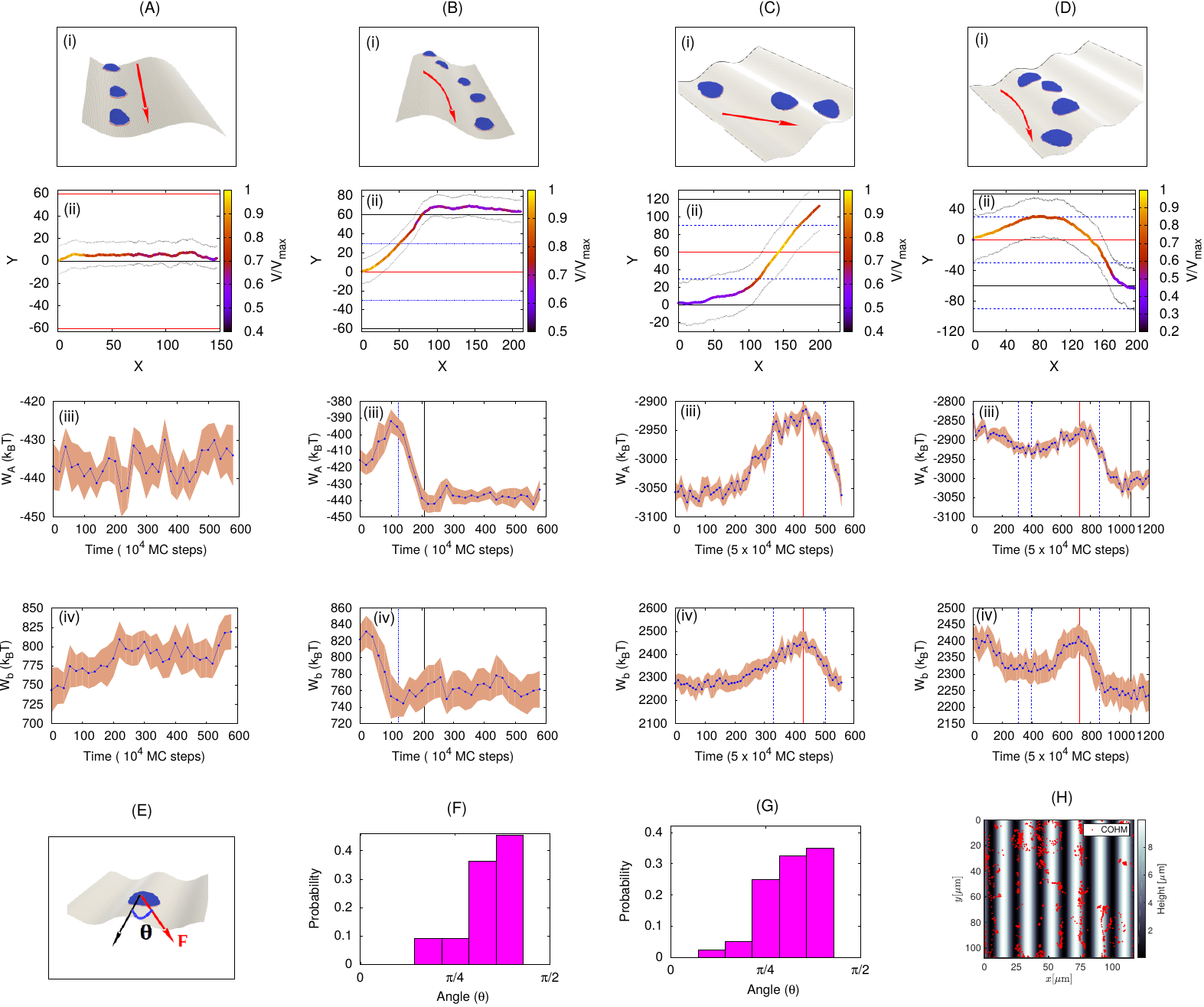}
\caption{Motile vesicle moving on a sinusoidal substrate with $y_m/R_{vesicle} \gg 1$. We use $z_m=10~l_{min};~y_m=120~l_{min}$ for sinusoidal substrate. (A) A small vesicle starting from the minimum of the sinusoidal substrate continues to migrate along its initial direction of migration. (B) Small vesicle starting from the maximum of the sinusoidal substrate shifts to the minimum of the substrate. (C) Large vesicle starting from the minimum of a sinusoidal  substrate (with $F=2.0 k_BT/l_{min}$), crosses the maximum and reaches the next minimum. (D) Large vesicle starting from the maximum of a sinusoidal substrate (with $F=1.0 k_BT/l_{min}$) initially tends to migrate along the positive $Y$-axis, then changes its direction of migration towards the negative $Y$-axis, and finally reaches the minimum. The panel (i) shows the snapshots (with red arrows showing the direction of migration), panel (ii) shows the trajectories, panel (iii) shows the adhesion energy with time and panel (iv) shows the bending energy with time. (E) We define migration angle ($\theta$) as the angle between the direction of migration of the vesicle (towards the net active force $F$) and the axis of the sinusoidal substrate ($x$-axis). (F) The distribution of angle at which the vesicle crosses the maxima, generated from simulation. Here we only use the data for large vesicle, as small vesicle in this case does not cross the ridges. (G) The distribution of angle at which the vesicle crosses the maxima, generated from the experimental trajectories of Ref. \cite{song2015sinusoidal}. (H) The accumulated positions of center-of-mass of \textit{Dictyostelium discoideum} cells over time.  For small vesicle (A-B), we use $N=607$, $E_{ad}=3.0~k_BT$, $F=4.0 k_BT/l_{min}$ and $\rho=4.9\%$. For large vesicle (C-D), we use $N=3127$, $E_{ad}=2.0$ and $\rho=2.4 \%$.}
\label{large_vesicle_sinusoidal}
\end{figure}

We start by studying the vesicle migration on a sinusoidal substrate where the sinusoidal wavelength is larger than the diameter of the adhered cell, and the vesicle can migrate while it is roughly in a region with only one type of substrate curvature over its entire contact surface with the substrate: the groove/ridge width $y_m/2$ is about twice larger than the cell diameter $2R_{vesicle}$. In Fig.\ref{large_vesicle_sinusoidal}(i,ii) we show the configurations and trajectories of a motile vesicle that was placed initially either on the bottom (Fig.\ref{large_vesicle_sinusoidal}A) or top (Fig.\ref{large_vesicle_sinusoidal}B) of the sinusoidal surface undulation. The vesicle is initially aligned parallel to the surface undulations (along the $x$-axis).

When we use a vesicle of small size (Fig.\ref{large_vesicle_sinusoidal}A,B) we find simple dynamics on the sinusoidal surface: when initiated inside the groove, it maintains its aligned direction of motion (Fig.\ref{large_vesicle_sinusoidal}A, Movie-{\textcolor{blue}{S2}}). When initiated on the ridge, the vesicle quickly reorients to almost perpendicular direction of motion, slides to the nearby groove, where it resumes its aligned migration (Fig.\ref{large_vesicle_sinusoidal}B, Movie-{\textcolor{blue}{S3}}). 

A larger vesicle (surface area 5 times larger) on the same sinusoidal substrates exhibits more complex dynamics (Fig.\ref{large_vesicle_sinusoidal}C,D; Movie-{\textcolor{blue}{S4}},{\textcolor{blue}{S5}}). This is due to the vesicle now extending over a larger surface and simultaneously spanning more of the two signs of the substrate curvatures. For example, when started in the groove (Fig.\ref{large_vesicle_sinusoidal}C(i,ii)), it is affected by the nearby ridge, which causes a reorientation similar to that observed in  Fig.\ref{large_vesicle_sinusoidal}B(i,ii). Occasionally, the larger vesicle remain aligned in the groove (see Fig.S4A), but its leading edge aggregate often breaks up, sometimes leading to a loss of the motile phenotype. When the larger vesicle is initiated on the ridges it reorients towards the nearby groove, but due to spanning both sides of the ridge, the vesicle can change its direction during this process (Fig.\ref{large_vesicle_sinusoidal}D(i,ii)). More examples of these dynamics are shown in SI Sec. S4, Fig.S4 (also see Movie-{\textcolor{blue}{S6}},{\textcolor{blue}{S7}},{\textcolor{blue}{S8}}). 

In order to understand this behaviour, we plot the adhesion ($W_A$, Eq.S6) and bending energies ($W_b$, Eq.S1) of the vesicle as it is moving between the ridge and groove regions (Fig.\ref{large_vesicle_sinusoidal}A-D(iii, iv)). We note that both the adhesion and bending energies are roughly constant when the vesicle migrates in the groove. When the vesicle shifts from the ridge to the groove, both the adhesion and bending energies decreases, driving the preference for the vesicle to remain inside the groove. This is easy to understand, as the vesicle can adhere more snugly when ``filling" the concave groove, with lower bending energy at the vesicle rim, compared to being more curved on the ridge region. On the other hand, the curved nucleators form stronger bonds between themselves ($W_d$, Eq.S4), and therefore a more robust leading edge cluster, when on the ridge (Fig.S5). However, the changes in this energy term are small compared to the changes in the bending and adhesion energy. These observations explain why energetically it is overall more favorable for the vesicle to reside in the grooves, while the more cohesive leading edge cluster gives rise to faster motility when the vesicle crosses the ridges. Note that the cell-substrate adhesion energy was previously identified as the driving mechanism for the tendency of cells to accumulate in concave grooves and pits \cite{vassaux2019biophysical}.

Our theoretical results shown in Fig.\ref{large_vesicle_sinusoidal}(A-D)(i,ii) are similar to the experimental observations of T lymphocytes migrating on sinusoidal surfaces \cite{song2015sinusoidal}. In these experiments it was found that cells mostly migrate inside, and aligned with the grooves, while occasionally crossing the ridges rapidly and at large angles. The simulations indicate that the vesicle tends to cross the ridges at large angles (Fig.\ref{large_vesicle_sinusoidal}F), as observed in experiments \cite{song2015sinusoidal} (Fig.\ref{large_vesicle_sinusoidal}G). A similar behavior was observed in migrating \textit{Dictyostelium discoideum} cells on a sinusoidal substrate, as shown in Fig.\ref{large_vesicle_sinusoidal}H \cite{Marcel-thesis}. The positions of the center-of-mass of the cells over time shows that this cell type also tends to stay within the grooves, and avoids the ridges. See sec. S2 for the details of experimental methods.

%%%%%%%%%%%%%%%%%%%%%%%%%%%%%%%%%%%%%%%%%%%%%%%%%%%%%%%%%%%%%%%%%%%%%%%%%%%%%%%%%%%%%%%%%%%%%

\subsection{Cellular migration on sinusoidal surfaces: small wavelength}

\begin{figure}[h!]
\centering
\includegraphics[scale=1.05]{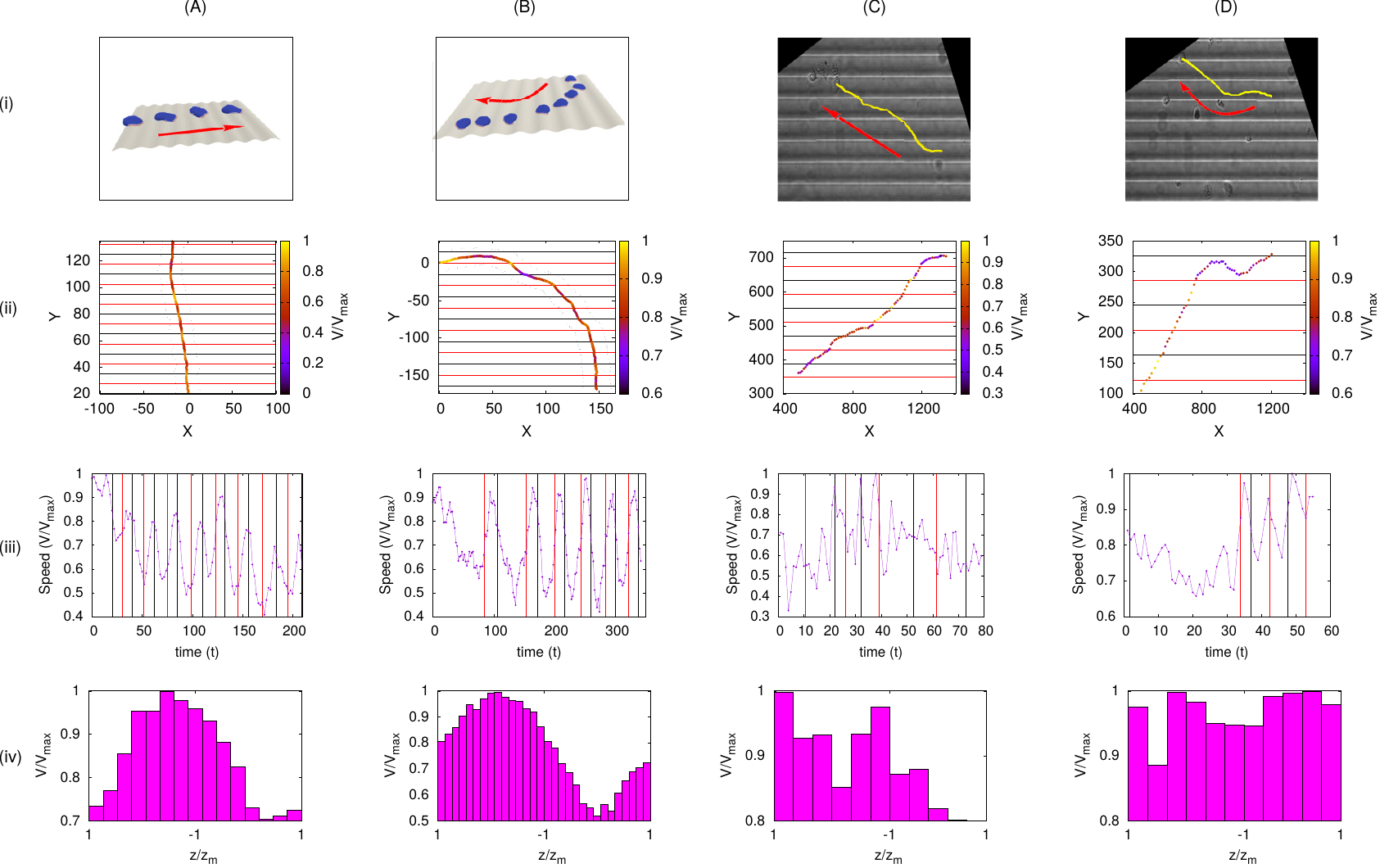}
\caption{Motile vesicle moving on a sinusoidal substrate with $y_m/R_{vesicle} \sim 1$. In this case, we only use small vesicle for simulation results. (A) A vesicle starting in the orthogonal direction of the sinusoidal substrate continues to migrate in the same direction. Here, we use $z_m=1~l_{min};~y_m=15~l_{min}$ for the sinusoidal substrate. (B) A vesicle started from the maximum of the sinusoidal substrate slowly changes its migration direction and becomes orthogonal to the sinusoidal axis. Here, we use $z_m=2~l_{min};~y_m=30~l_{min}$ for the sinusoidal substrate. (C-D) Migrating \textit{karatocytes} on sinusoidal substrate. Here, (i) shows the snapshots (red arrows are showing the direction of migration), (ii) shows the trajectories, and (iii) shows the variation of speed of the vesicle/cell with time. For simulation results, we use $N=607$, $E_{ad}=3.0~k_BT$, $F=4.0 k_BT/l_{min}$ and $\rho=4.9\%$.}
\label{small_vesicle_sinusoidal}
\end{figure}

Next we consider the case where the vesicle radius and the wavelength of the sinusoidal undulations are of the same order, so that a vesicle spans both the ridge and the nearby groove(s). Here, we use sinusoidal variations of two types: $z_m=1;~y_m=15$ and $z_m=2;~y_m=30$ keeping the ratio of $z_m/y_m$ fixed.

We find that when we start with a vesicle that is either orthogonal (Fig.\ref{small_vesicle_sinusoidal}A, Movie-{\textcolor{blue}{S9}})  or parallel (Fig.\ref{small_vesicle_sinusoidal}B, Movie-{\textcolor{blue}{S10}}) to the sinusoidal pattern, the vesicle eventually settles to migrate mostly in the orthogonal direction. However, the speed of the vesicle shows clear oscillatory behaviour (Fig.\ref{small_vesicle_sinusoidal}A,B(iii)). When the vesicle travels from ridge to groove, it moves towards lower energies and thereby moves faster, while in the opposite case, it slows down. These speeds seem to be periodic along the orthogonal direction to the grooves and ridges of the sinusoidal pattern. In Fig.S6(A-B) we show the average speed of the migrating vesicle at different positions of its center-of-mass between two maxima of the sinusoidal pattern, showing clear periodicity.

Note that when the vesicle was aligned inside the groove (initial condition in Fig.\ref{small_vesicle_sinusoidal}B), it shows some tendency to persist inside the groove. This tendency gives rise to staircase-like trajectories when the vesicle moves at some oblique angle with respect to the sinusoidal pattern (Fig.\ref{small_vesicle_sinusoidal}B(ii)). We show more simulations of this type in Fig.S7 (also see Movie-{\textcolor{blue}{S11}},{\textcolor{blue}{S12}},{\textcolor{blue}{S13}},{\textcolor{blue}{S14}}).

We compare these simulations to experiments using fish keratocytes migrating on sinusoidal substrates \cite{mohammed2019substrate}, with similar ratio of cell size and sinusoidal wavelength (Movie-{\textcolor{blue}{S15}},{\textcolor{blue}{S16}}). Fish keratocytes is a perfect cellular system to be compared to vesicles since they are persistent and polarized cells that contain a large lamellipodium driven by protrusive forces exerted by actin polymerization. In Fig.\ref{small_vesicle_sinusoidal}C,D, we show two typical trajectories, where the cell migrates in a staircase-like trajectory (Fig.\ref{small_vesicle_sinusoidal}C(i,ii)) or leaves the groove and moves orthogonal to the pattern (Fig.\ref{small_vesicle_sinusoidal}D(i,ii); Fig.S8(B-D)). The speed of the cell shows similar oscillatory behaviour as observed in our simulation (Fig. \ref{small_vesicle_sinusoidal}C,D(iii)). However, due to the noisy cell speed extracted from the experimental trajectories, we could not identify a clear relation between the mean speed and the cell position within the sinusoidal pattern (Fig.S6(C-D)). The experimental speed can be affected by stick-slip cellular retractions and inhomogeneities in the cell-substrate adhesion, which are absent in the simulations. In Fig.S8 we show more experimental trajectories of migrating keratocytes on the sinusoidal substrate, similar to  Fig.\ref{small_vesicle_sinusoidal}C,D (also see Movie-{\textcolor{blue}{S17}},{\textcolor{blue}{S18}},{\textcolor{blue}{S19}},{\textcolor{blue}{S20}}).

Despite the favorable comparisons between the model and the experiments on sinusoidal surfaces, it is not easy to interpret the details of the migration process on these surfaces since they contain curvatures of opposite signs. We next explore the migration pattern of our model vesicle, and living cells, on simpler curved surfaces of uniform curvature.

%%%%%%%%%%%%%%%%%%%%%%%%%%%%%%%%%%%%%%%%%%%%%%%%%%%%%%%%%%%%%%%%%%%%%%%%%%%%%%%%%%%%%%%%%%%%%
%%%%%%%%%%%%%%%%%%%%%%%%%%%%%%%%%%%%%%%%%%%%%%%%%%%%%%%%%%%%%%%%%%%%%%%%%%%%%%%%%%%%%%%%%%%%%%
\subsection{Migration outside and inside a cylindrical surface (fiber and tube)}
\begin{figure}[h!]
\centering
\includegraphics[scale=1.5]{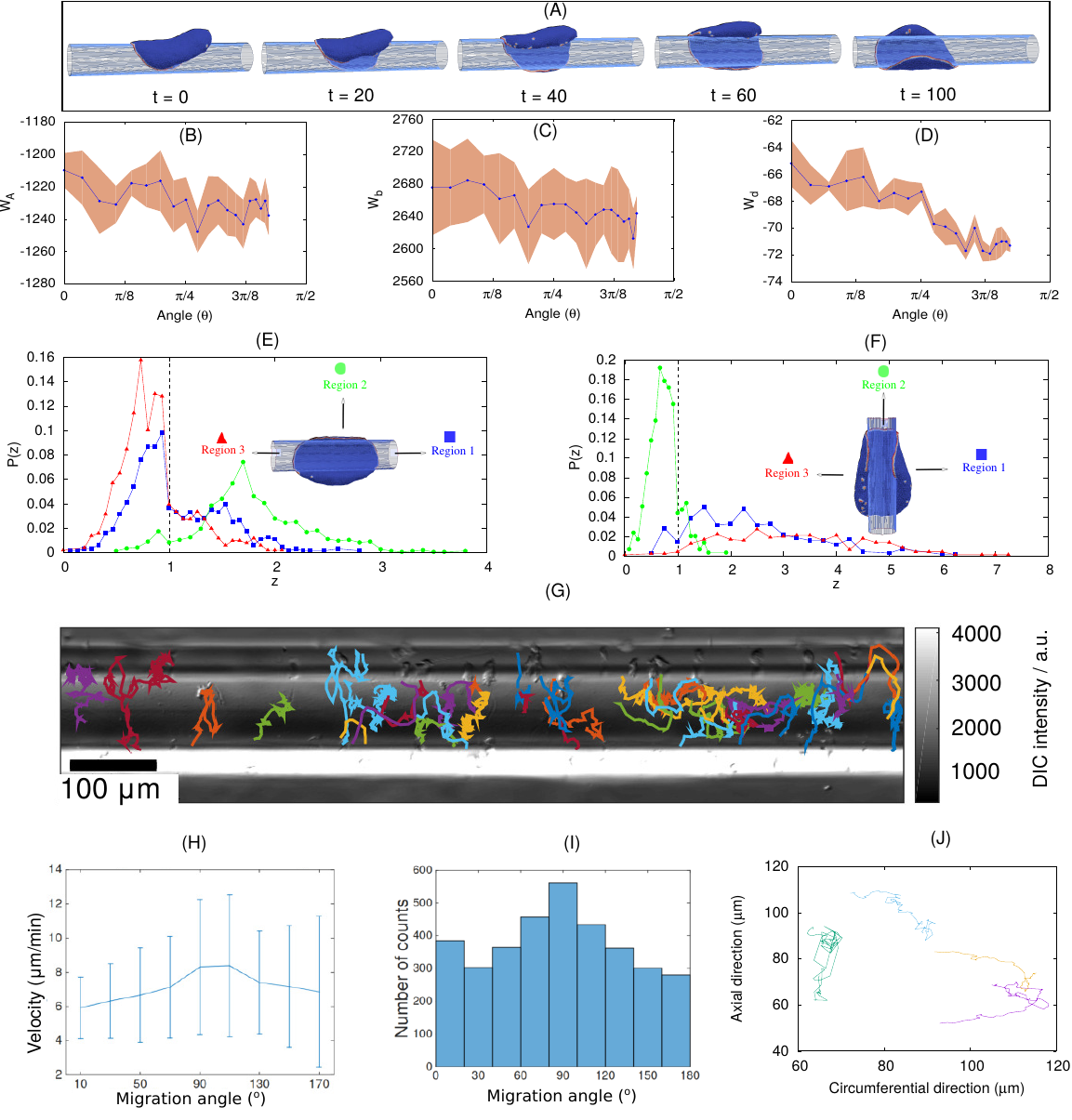}
\caption{Vesicle migrating outside of a cylindrical fiber. (A) Configurations of the motile vesicle migrating on a fiber, initially in the axial direction, and finally reorients to rotate circumferentially. (B) The adhesion, (C) bending, and (D) binding energies of the vesicle as function of its migration angle during the reorientation process shown in (A). (E) Distribution of the distance $z$ of a curved protein along the leading edge from the cylindrical surface, when the vesicle is oriented circumferentially. We plot this distance distribution for three different sections of the leading edge, along three directions, as defined in the inset. The part of the distributions that are on the left side of the vertical dashed line (at $z=1$) represent adhered proteins. (F) Same as (E), when the vesicle is oriented axially. (G) Trajectories of different D.d. cells on the fiber  of diameter $160~ \mu m$ \cite{blum2015curvotaxis}. (H) The distribution of migration speeds of D.d. cells, as function of its migration angle.  (I) Distribution of migration angles of D.d. cells on the fiber. (J) The trajectories of MDCK cells migrating on a fibers of $50-70~\mu m$ in diameter. For simulation results, we use $R=10~l_{min}$, $F=2.0$, $E_{ad}=1.0$ and $\rho=2.4 \%$.}
\label{vesicle_on_cylinder}
\end{figure}

In order to gain a deeper understanding of the curvature-dependent motility in our model, we simulate the vesicle motion on a surface of uniform curvature, such as the convex curvature of the external surface of a cylinder (fiber). In Fig.\ref{vesicle_on_cylinder}A we plot the dynamics of a motile vesicle, when initially it was aligned along the axis of the fiber (of radius $R=10~l_{min}$). We find that the vesicle spontaneously shifts its orientation, and ends up rotating along the circumferential direction, as the final steady-state of the system (Fig.\ref{vesicle_on_cylinder}A, Movie-{\textcolor{blue}{S21}}). This tendency, to polarize and migrate perpendicular to the axis of the fiber, explains naturally the tendency of the vesicle to migrate perpendicular to the undulation pattern, when moving over the ridges of the sinusoidal surfaces (Figs.\ref{large_vesicle_sinusoidal},\ref{small_vesicle_sinusoidal}).

We can understand the driving force for this re-orientation of the migration, by plotting the adhesion, bending and protein-binding energies of the vesicle during this process (Fig.\ref{vesicle_on_cylinder}B-D) as a function of the migration angle, which is defined as the angle between the direction of motion and the fiber axis (see SI Sec. S9,S10; Fig.S9,S10 for more details). We see that there is a small gain in adhesion (decrease in adhesion energy), decrease in overall bending energy, and a small decrease in the protein binding energy. When oriented circumferentially, the leading edge active forces can stretch the vesicle sideways along the cylinder's axis, which is efficient in increasing the adhered area along a direction of low curvature, by keeping the membrane close to the fiber surface (Fig.\ref{vesicle_on_cylinder}E). By comparison, when the vesicle is oriented along the axis (Fig.\ref{vesicle_on_cylinder}F) only a small region of the leading edge, along the axis, can pull the membrane close to the fiber and maintain its adhesion. The parts of the leading edge that point along the circumferential direction are less effective in increasing the adhered area due to pulling the membrane off the surface, as well as increasing its bending energy.

This predicted tendency for cells to rotate around fibers, when their migration is driven by a lamellipodial protrusion, is nicely verified by experimental data on \textit{Dictyostelium discoideum} cells \cite{blum2015curvotaxis}. The observed trajectories of migration are biased along the circumferential direction (Fig.\ref{vesicle_on_cylinder}G, Movie-{\textcolor{blue}{S22}}), as shown by the peak in the distribution of cellular migration direction (Fig. \ref{vesicle_on_cylinder}I). The speed of the cells  was also found to be maximal along the circumferential direction (Fig. \ref{vesicle_on_cylinder}H).

Furthermore, it was found experimentally that the tendency of the cells to migrate circumferentially decreased as the fiber radius increased (Fig.S11) \cite{blum2015curvotaxis}. Our model can offer an explanation of this trend, as we find that the energetic advantage of the circumferential orientation in our simulations decreases with increasing fiber radius.

A similar tendency was observed for motile MDCK cells on a fiber (Movie-{\textcolor{blue}{S23}}). The trajectories in Fig. \ref{vesicle_on_cylinder}J show that these cells were either migrating in fast and highly directed bursts along the circumferential direction (Fig.S12), or moving slowly in random motion along the axial direction. This agrees with the model's prediction that the lamellipodia's leading edge is more robust along the circumferential direction, which should result in more persistent motion in this direction. 

Previous studies with MDCK cells moving on very thin fibers (fiber cross-section circumference same or smaller than the cell diameter) reported a bi-phasic migration pattern \cite{yevick2015architecture}. Isolated cells were sometimes observed to migrate axially with high speed, and with a very small adhered surface area. The cell body in these cases exhibits a highly rounded shape, typical of cells under strong contractile forces. Such contractile forces are outside the present model, and we therefore do not expect to reproduce this axially motile phenotype \cite{guetta2015protrusive}. However, a second phenotype was observed in these experiments, when cells spread and adhere strongly to the fiber surface. During these times the cells seem to exhibit short rotation periods around the fiber circumference \cite{yevick2015architecture}, but their duration were too short to be conclusive. In addition, the overall orientation of the actin fibers in a confluent monolayer of cells on the fiber was found to be circumferential, in agreement with the orientation of the isolated cell in our simulations.

\begin{figure}[h!]
\centering
\includegraphics[scale=0.32]{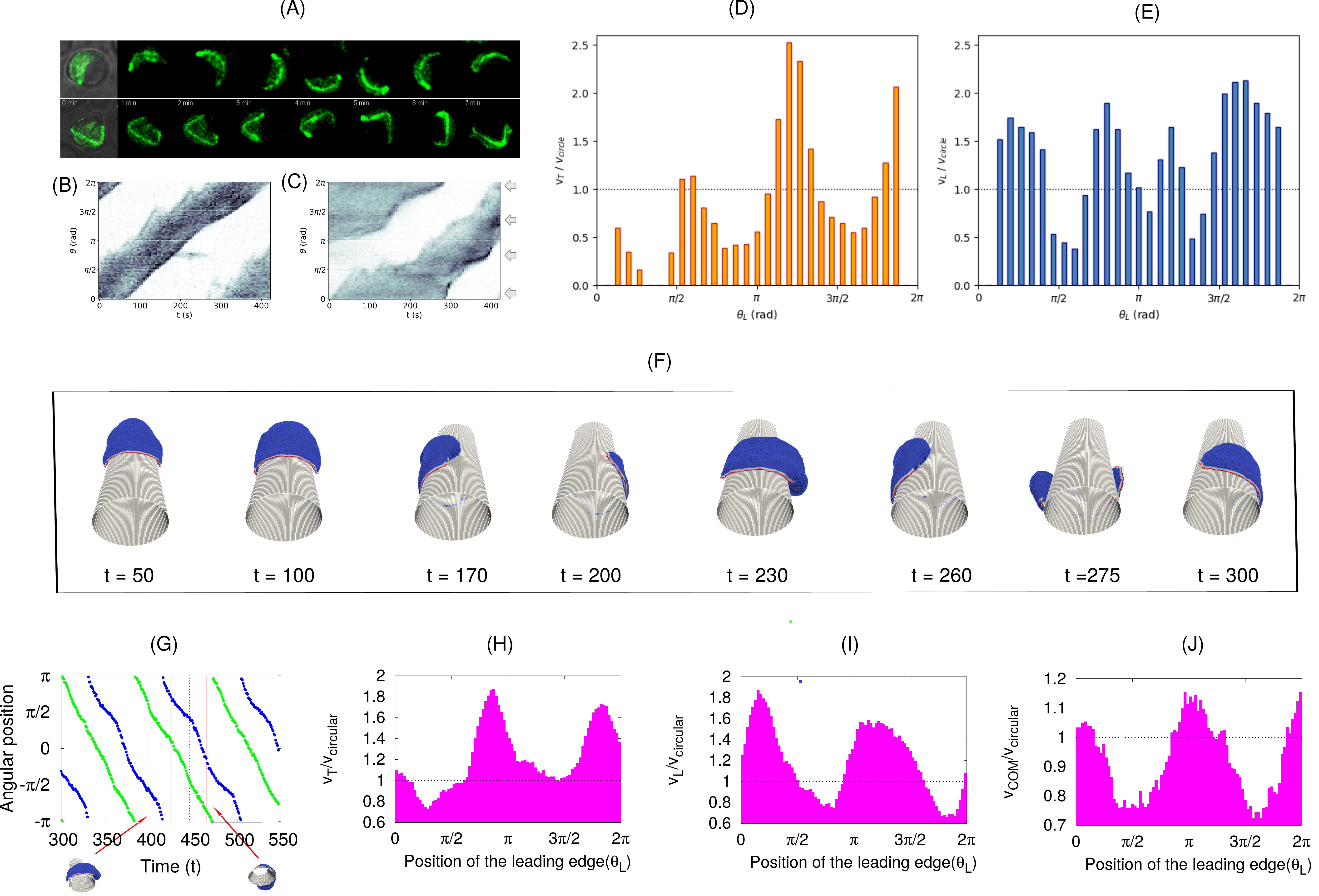}
\caption{Curvature sensing on micropillars. (A) Timelapse snapshots of \textit{Dictyostelium discoideum} cells moving along the surface of a circular (top) or triangular (bottom) shaped micropillar. In both cases, the field of view shown is 30 x 30 um. Cells express LifeAct-GFP. (B) Kymograph of the actin signal for the cell shown in (A) on the round micropillar. The intensity has been integrated in each angular slice and colour coded from white (0) to black (max). (C) Same as (B) for the triangular shaped pillar. The arrows indicate the positions where the triangle corners are located. (D) Variation in the speed of the trailing edge $V_T$ (scaled by the average speed for the fiber with circular cross-section) as a function of the position of the leading edge ($\theta_L$). (E) Variation in the speed of the leading edge ($V_L$) as a function of the position of the leading edge ($\theta_L$). (F) Snapshots of the migrating vesicle on a fiber with elliptical cross-section with aspect ratio $r=1.54$. The vesicle initially migrates in the axial direction and finally reorients along the circumferential direction for longer time $(t>200)$. (G) Kymograph of the leading edge (blue circles) and the trailing edge (green circles) when the vesicle is rotating circumferentially $(t>300)$. (H) Speed of the trailing edge of the vesicle $V_T$ (scaled by the speed for the fiber with circular cross-section $V_{circular}$) as a function of the position of the leading edge ($\theta_L$) over a full period. (I) Speed of the leading edge of the vesicle $V_L$ as a function of the position of the leading edge ($\theta_L$) over a full period. (J) Speed of the centre of mass (com) of the vesicle $V_{COM}$ as a function of the position of the leading edge ($\theta_L$) over a full period.  For simulation, we use $R_x=12~l_{min}$, $R_y=7.773~l_{min}$, $E_{ad}=1.5~k_BT$, $F=2.0~k_BT/l_{min}$,  and $\rho=2.4~\%$.}
\label{pillars}
\end{figure}

Another example for spontaneous rotational migration of cells on cylindrical surfaces, is shown in Fig.\ref{pillars}(A-C) (Movie-{\textcolor{blue}{S24}},{\textcolor{blue}{S25}}). Here \textit{Dictyostelium discoideum} cells are shown to rotate persistently on the external surface of pillars with circular cross-section. On pillars with triangular cross-section, we find that the cells slow down periodically whenever they cross the higher curvature corners (\ref{pillars}(D,E)).

Note that when cells are migrating on extremely thin fibers, the motility mode is very different, driven by elongated and thin protrusions on either side of the cell \cite{guetta2015protrusive}. There is no single lamellipodium that drives the migration, and no global rotation of the cell around the fiber. However, the leading edges of the protrusions tend to coil around the fiber. We suggest that this behavior is driven by the same mechanism that we identified here to cause global rotations on larger fibers \cite{sadhu_coiling}.

In order to verify the above predictions, we simulated the migration of the motile vesicle on a fiber of elliptical cross-section, such that the rotating vesicle experiences different curvatures periodically (Fig.\ref{pillars}F, Movie-{\textcolor{blue}{S26}}). By plotting the kymograph (Fig.\ref{pillars}G) and the speed as function of position (Fig.\ref{pillars}H-J), we find periodic variations in the speed of the vesicle that are similar to those observed in the experiments (Fig.\ref{pillars}B-E). Note that the experiments exhibit three peaks, due to the triangular shape, compared to two peaks in the simulations on the elliptic cross-section.

In the SI Sec. S13, Fig.S13, we compare the dynamics of migrating vesicles on elliptical fibers of different aspect ratio $r=R_x/R_y$ (Movie-{\textcolor{blue}{S27}},{\textcolor{blue}{S28}}). We note that it takes more time for the vesicle to reorient towards the circumferential direction as the aspect ratio $r$ increases (Fig.S13 A,B). For the largest aspect ratio that we tested ($r=2.87$), the vesicle does not reorient at all (Fig.S13 C), due to the sharp corners that present bending energy barriers. This inhibition of rotation over the sharp corners is similar to the inhibition of coiling at the leading edge of cellular protrusions, calculated and observed when cells spread over fibers \cite{sadhu_coiling}.

%PILLARS EXPERIMENTS OF CARSTEN BETA

Finally, we simulated the migration of the motile vesicle inside a cylindrical tube (see SI sec. S14, Fig.S14,S15, Movie-{\textcolor{blue}{S29}},{\textcolor{blue}{S30}}). We find that the vesicle prefers to migrate along the tube axis, but easily loses its motility, and can even form ``bridges" across the tube axis. This prediction was verified experimentally for MDCK cells, which were found to be weakly motile, with the most persistent motility periods aligned with the tube axis or spread across the tube axis (Fig.S15G, Movie-{\textcolor{blue}{S31}}).

%%%%%%%%%%%%%%%%%%%%%%%%%%%%%%%%%%%%%%%%%%%%%%%%%%%%%%%%%%%%%%%%%%%%%%%%%%%%%%%%%%%%%%%%%%%%%%%%%%%%
\section{Conclusion}

We demonstrated here that a minimal physical model of a motile cell, based on very few ingredients and energy terms, is able to describe and explain several curvotaxis features of lammelipodia-based cell migration on curved adhesive substrates. Within this minimal model, cell migration arises when a leading edge cluster of highly curved membrane protein complexes forms, due to these protein complexes being both highly curved, binding to each other and exerting protrusive forces on the membrane. These forces represent in the model the pressure exerted on the membrane when actin polymerization is initiated at the membrane by these curved protein complexes, which contain actin nucleation factors (such as WAVE \cite{TakenawaWave,PollittWave,StradalWave}). The curvotaxis features that the model explains, such as the tendency of motile cells to migrate aligned within grooves, avoid ridges, and rotate around fibers, all arise due to minimization of the adhesion and bending energies of the vesicle. The advantage of simple, physical models is demonstrated here, exposing general mechanisms that are universal and not cell-type-specific.

The curvotaxis property of the motile “minimal-cell” is shown to be a truly emergent phenomenon of the whole motile vesicle. Within our model, the energy minimization that aligns the migration of the ``minimal-cell" arises from shape changes of the whole vesicle in response to the imposed curved surface and the organization of the curved membrane complexes that form the leading-edge cluster (and motility). The curved membrane complexes are sensitive to curvature on a much smaller length-scale compared to the cell-size, and therefore do not directly determine the preferred curvotaxis response of the whole motile vesicle.

Eukariotic cells contain numerous additional components that our simple model does not contain, such as the effects of contractility, stress-fibers and internal organelles (such as the big nucleus), which can all affect migration on curved substrates. Nevertheless, the agreement between the predictions of the model and the observations of curvotaxis in different types of motile cells, suggests that these simple energetic considerations may drive curvotactic features in cells, despite the biochemical complexity and differences between cells. These results demonstrate that complex cellular behavior may have physical underpinnings, with added layers of biological complexity and regulation. The framework presented here could serve in the future to explore cell migration in more complex geometries \cite{pieuchot2018curvotaxis,fischer2021contractility}, and over soft substrates (such as other cells) with dynamic curvature.

%%%%%%%%%%%%%%%%%%%%%%%%%%%%%%%%%%%%%%%%%%%%%%%%%%%%%%%%%%%%%%%%%%%%%%%%%%%%%%%%%%%%%%%%%%%%%%%%%%%%%%%%
%%%%%%%%%%%%%%%%%%%%%%%%%%%%%%%%%%%%%%%%%%%%%%%%%%%%%%%%%%%%%%%%%%%%%%%%%%%%%%%%%%%%%%%%%%%%%%%%%%%%
\section{Author contributions}
R.K.S.,  S.P., A.I. and N.S.G. developed the theoretical model; R.K.S. and N.S.G. conceived, designed and implemented the analysis of the model, and prepared the manuscript; M.L. and S.G. conceived and performed the experiments on keratocytes migrating on sinusoidal substrate and contributed the data; W.X. and B.L. performed the experiments and analysis of the MDCK cells migrating inside the tube and outside of fiber and contributed the data; M.S., Christoph Blum, M.T., O.S. and E.B. conceived and supervised the experiments on D.d. cells migrating on fiber and sinusoidal substrate and contributed the data; Carsten Beta and C.M.T. conceived and supervised the experiments on D.d. cells migrating on pillars and contributed the data; All authors reviewed and edited the manuscript.
%%%%%%%%%%%%%%%%%%%%%%%%%%%%%%%%%%%%%%%%%%%%%%%%%%%%%%%%%%%%%%%%%%%%%%%%%%%%%%%%%%%%%%%%%%%%%%%%%%%%
\section{Acknowledgment}
We thank Junsang Doh  and collaborators for providing the data for the migration of T lymphocytes of sinusoidal wavy surfaces. N.S.G. is the incumbent of the Lee and William Abramowitz Professorial Chair of Biophysics, and acknowledges support by the Ben May Center for Theory and Computation, and the Israel Science Foundation (Grant No. 207/22). This research is made possible in part by the historic generosity of the Harold Perlman Family. R.K.S. acknowledges the support from ANR (ANR-19-CE11-0002-03). A.I. and S.P. were supported by the Slovenian Research Agency (ARRS) through the Grant No. J3-3066 and J2-4447 and Programme No. P2-0232. This work was supported by the Marie Skłodowska-Curie Actions, Individual Fellowship, Project: 846449 (to W.X.) and the ``Initiatives d’excellence" (Idex ANR-11-IDEX-0005-02) transverse project BioMechanOE (TP5) (to W.X.). This work was supported by the European Research Council (Grant No. Adv-101019835 to B.L.), LABEX Who Am I? (ANR-11-LABX-0071 to B.L. and W.X.) and the Ligue Contre le Cancer (Equipe labellisée 2019 to B.L. and W.X.) and the ANR PRC LUCELL grant (ANR-19-CE13-0014-01 to B.L. and W.X.), and DIM-ELICIT 2019: Equipment support, Région Ile-de-France (to W.X. and B.L.). B.L. and W.X. acknowledge the ImagoSeine core facility of the IJM, member of IBiSA and France-BioImaging (ANR-10-INBS-04) infrastructures. The research of Carsten Beta and C.M.T. has been partially funded by the Deutsche Forschungsgemeinschaft (DFG), Project-ID No. 318763901—SFB1294. S.G. acknowledges funding from FEDER Prostem Research Project no. 1510614 (Wallonia DG06), the F.R.S.-FNRS Epiforce Project no. T.0092.21, the F.R.S.-FNRS Cellsqueezer Project no. J.0061.23, the F.R.S.-FNRS Optopattern Project no. U.NO26.22 and the Interreg MAT(T)ISSE project, which is financially supported by Interreg France-Wallonie-Vlaanderen (Fonds Européen de Développement Régional, FEDER-ERDF). M.L. is financially supported by the WBI-World Excellence Grant Programme for long-term scholarship.

%%%%%%%%%%%%%%%%%%%%%%%%%%%%%%%%%%%%%%%%%%%%%%%%%%%%%%%%%%%%%%%%%%%%%%%%%%%
%%%%%%%%%%%%%%%%%%%%%%%%%%%%%%%%%%%%%%%%%%%%%%%%%%%%%%%%%%%%%%%%%%%%%%%%%%%
%%%%%%%%%%%%%%%%%%%%%%%%%%%%%%%%%%%%%%%%%%%%%%%%%%%%%%%%%%%%%%%%%%%%%%%%%%
\pagebreak

\renewcommand{\thefigure}{S-\arabic{figure}}
\renewcommand{\thesection}{S-\arabic{section}}
\renewcommand{\theequation}{S-\arabic{equation}}
\setcounter{equation}{0}
\setcounter{section}{0}
\setcounter{figure}{0}
%\renewcommand{\thetitle}{title}
%\renewcommand{\theauthor}{author}

%%%%%%%%%%%%%%%%%%%%%%%%%%%%%%%%%%%%%%%%%%%%%%%%%%%%%%%%%%%%%%%%%%%%%%%%%%
%%%%%%%%%%%%%%%%%%%%%%%%%%%%%%%%%%%%%%%%%%%%%%%%%%%%%%%%%%%%%%%%%%%%%%%%%%
%%%%%%%%%%%%%%%%%%%%%%%%%%%%%%%%%%%%%%%%%%%%%%%%%%%%%%%%%%%%%%%%%%%%%%%%%
\section*{{\Large{Supplementary Information for}} \\ {\large {A minimal physical model for curvotaxis driven by curved protein complexes at the cell's leading edge}}}

\author{$^1$Raj Kumar Sadhu\footnote{raj-kumar.sadhu@curie.fr} \footnote{Present address: Institut Curie, PSL Research University, CNRS, UMR 168, Paris, France}}
\author{$^2$Marine Luciano}
\author{$^3$Wang Xi}
\author{$^{4}$Cristina Martinez-Torres}
\author{$^5$Marcel Schr{\"{o}}der}
\author{$^5$Christoph Blum}
\author{$^5$Marco Tarantola}
%\author{$^6$Vahid Nasiri}
\author{$^6$Samo Peni\v{c}}
\author{$^6$Ale\v{s} Igli\v{c}}
\author{$^{4}$Carsten Beta}
\author{$^7$Oliver Steinbock}
\author{$^5$Eberhard Bodenschatz}
\author{$^3$Beno{\^{i}}t Ladoux}
\author{$^8$ Sylvain Gabriele}
\author{$^1$Nir S. Gov\footnote{nir.gov@weizmann.ac.il}}
\affiliation{$^1$Department of Chemical and Biological Physics, Weizmann Institute of Science, Rehovot 7610001, Israel.}
\affiliation{$^2$Department of Biochemistry, University of Geneva, Geneva Switzerland.}
\affiliation{$^3$ Universite Paris Cite, CNRS, Institut Jacques Monod, F-75013 Paris, France.}
\affiliation{$^{4}$Institute of Physics and Astronomy, University of Potsdam, Potsdam 14476, Germany.}
\affiliation{$^5$Max Planck Institute for Dynamics and Self-Organization, G{\"{o}}ttingen, Germany.}
\affiliation{$^6$Laboratory of Physics, Faculty of Electrical Engineering, University of Ljubljana, Ljubljana, Slovenia.}
\affiliation{$^7$Florida State University, Department of Chemistry and Biochemistry, Tallahassee, FL 32306-4390, USA}
\affiliation{$^8$Mechanobiology \& Biomaterials group, Interfaces and Complex Fluids Laboratory, Research Institute for Biosciences, CIRMAP, University of Mons, Mons Belgium.}

%\maketitle

\textbf{This PDF file includes:}\\

Supplementary Text\\

Secs. S1 to S14\\

Figs. S1 to S16\\

Legends for Movies S1 to S31\\

References (1-30) \\

\textbf{Other Supplementary Materials for this manuscript include the following:} \\

Movies S1 to S31
%%%%%%%%%%%%%%%%%%%%%%%%%%%%%%%%%%%%%%%%%%%%%%%%%%%%%%%%%%%%%%%%%%%%%%%%%%%%%%%%%%%%%
%%%%%%%%%%%%%%%%%%%%%%%%%%%%%%%%%%%%%%%%%%%%%%%%%%%%%%%%%%%%%%%%%%%%%%%%%%%%%%%%%%%%%%%%%%%%%%%%%%%%%%%%
\section{Theoretical Model}
\begin{figure}[h!]
\centering
\includegraphics[scale=0.92]{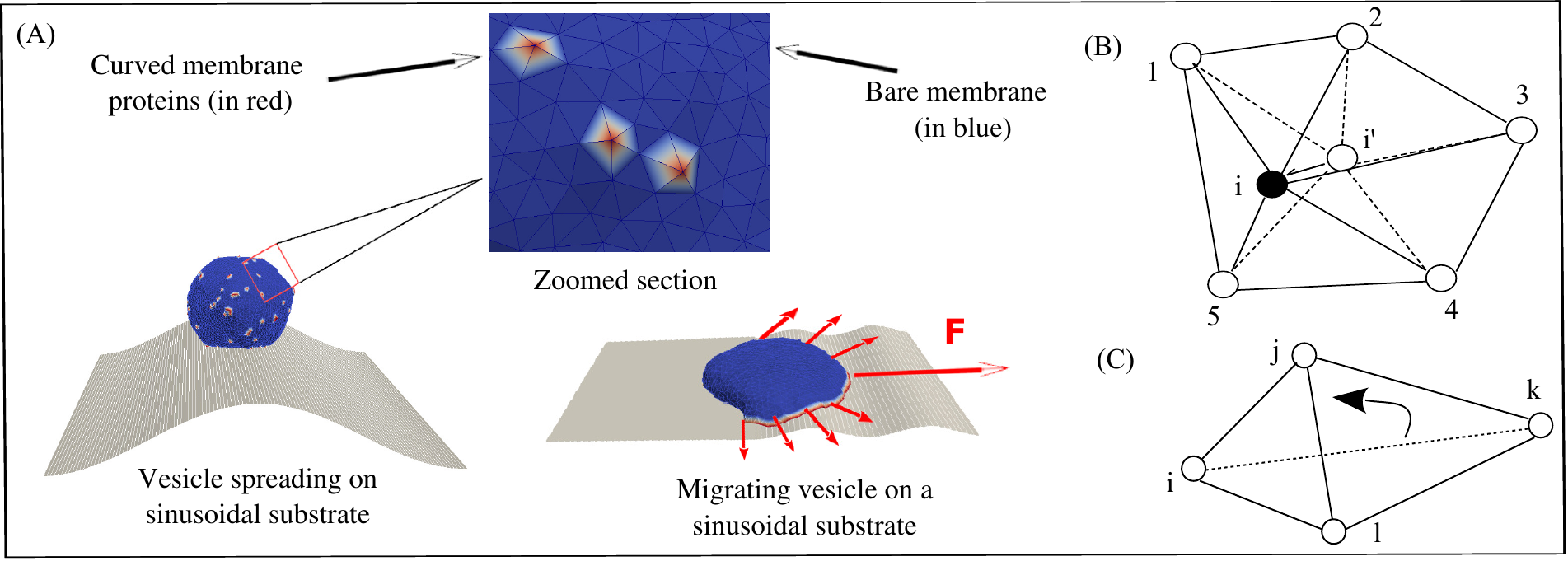}
\caption{Schematic representation of our model. (A) The vesicle is formed by a closed triangulated surface, having $N$ vertices connected to its neighbours with bonds. These bonds can change their length such that they are never below $l_{min}$, or above $l_{max}=1.7 l_{min}$. The red dots on the surface of the vesicle represents the curved membrane protein complexes with positive intrinsic curvature (convex), while the blue part represents bare membrane. A zoomed version of a small section of the vesicle surface is shown in the inset. We show here two possible initial conditions: (left) The vesicle starts with a spherical-like shape, adheres, spreads and migrates on the curved surfaces. (right) We generate a motile (crescent shaped) vesicle on a flat substrate and then deform the surface and let the vesicle evolve to conform to the deformed (curved) shape, and then allow it to migrate. (B) Vertex movement: The vertex $i'$ is moved to $i$. (C) Bond flip: The bond $i-k$ is flipped to bond $j-l$.}
\label{model-fig-SI}
\end{figure}

The migrating cell is represented in our theoretical model using a three-dimensional membrane vesicle. The vesicle is described by a closed triangulated surface having $N$ vertices, connected to their neighbours with bonds, and forming a dynamically triangulated, self-avoiding network, with the topology of a sphere \cite{RAMAKRISHNAN2014,Sadhu2021,miha2019,sadhu_coiling,Sadhu_phagocytosis,GOV-chapter25, DRAB-chapter26} (Fig.\ref{model-fig-SI}). The nodes that compose the vesicle surface can either represent the bare membrane (blue in Fig.\ref{model-fig-SI}), or represent membrane protein complexes with convex spontaneous curvature \cite{Luka2021}, that diffuse on the membrane surface, having nearest-neighbor attractive interaction with each other (red in Fig.\ref{model-fig-SI}). Convex protein or membrane curvature stands for a node that is locally protruding outwards, with respect to the vesicle interior.

We consider that each curved protein complex recruits actin polymerization, which gives rise to a local protrusive force that pushed the membrane. This is represented in our model as an active force ($F$) exerted at the site of the curved protein on the membrane, in the direction of the local outward normal to the vesicle surface. The simplifying assumption is that the actin polymerization that occurs near the membrane can be treated as a local force exerted directly at the site of the curved protein complex which includes actin nucleation factors (such as the WAVE complex \cite{TakenawaWave,PollittWave,StradalWave}).

The vesicle energy has therefore the following contributions: The continuum version of the bending energy, is given by,

\begin{equation}
    W_b=\frac{\kappa}{2} \int_A (H - H_0)^2 dA,
    \label{bendingE}
\end{equation}

where $\kappa$ is the bending rigidity, $H$ is the mean local curvature of the membrane surface, $H_0$ is the local spontaneous curvature, and the integral is over the entire surface. The vesicle contains curvature sensitive protein complexes, that occupy vertices with an overall density $\rho=N_c/N$, where $N_c$ is the number of such protein nodes, and $N$ is the total number of vertices on the vesicle. These protein nodes have a positive (convex) spontaneous curvature ($H_0>0$), while the bare membrane nodes have zero spontaneous curvature. In our simulations we use a discrete version of the bending energy \cite{Helfrich1973,ESPRIU1987271} which we calculate in the following way:

In the absence of any spontaneous curvature, the integration of the square of the mean curvature over the entire surface can be discretized in the following way \cite{ESPRIU1987271}:

$$ \int_A H^2 dA = \sum _i \frac{1}{\sigma_i} \left[ \sum_{j(i)} \frac{\sigma_{ij}}{d_{ij}} (\textbf{R}_i - \textbf{R}_j)\right]^2$$

where the outer sum runs over all the vertices and the inner sum runs over all the neighbours of $i$-th vertex. $\textbf{R}_i$ is the radial vector of the vertex $i$, $d_{ij}$ is the distance between the vertices $i$ and $j$, $\sigma_i$ is the area of the cell (formed by the vertex $i$ and all its neighbours) in the dual lattice defined as,

$$\sigma_i = \frac{1}{4} \sum_{j(i)} \sigma_{ij} d_{ij}$$

where $\sigma_{ij}$ is the distance between the vertices $i$ and $j$ in the dual lattice.

In the presence of curved membrane proteins, the spontaneous curvature of the $i$-the vertex is $c_i$ (say). Then the bending energy of the $i$-th vertex can be written as,

\begin{equation}
W_b(i) = \frac{\kappa}{2} \sigma_i \left(\frac{h_i}{\sigma_i} - c_i \right)^2
\end{equation}

where,

$$h_i^2 = \left[ \sum_{j(i)} \frac{\sigma_{ij}}{d_{ij}} (\textbf{R}_i - \textbf{R}_j)\right]^2$$

Thus, the total bending energy of the vesicle can be written as,

\begin{equation}
    W_b = \sum_i W_b(i)
\end{equation}

where, the sum runs over all the vertices of the vesicle.

The direct binding energy between the protein complexes on nearest-neighbour nodes is given by, 

\begin{equation}
    W_d = -w\sum_{i<j} {\cal H} (r_0 - r_{ij}),
    \label{eq:bind}
\end{equation}

where, ${\cal H}$ is the Heaviside step function, $r_{ij}=|\overrightarrow{r}_j - \overrightarrow{r}_i|$ is the distance between proteins, $\overrightarrow{r}_i, \overrightarrow{r}_j$ are the position vectors for $i,j-th$ proteins, and $r_0$ is the range of attraction, $w$ is the strength of attraction. The range of attraction is chosen such that only the proteins that are in neighbouring vertices can bind to each other. 

These curved protein complexes also recruit actin filaments that polymerize at the location of these proteins. We assume that the direction of these forces are normally outward of the local surface containing the proteins. The active energy is given by, 

\begin{equation}
   \Delta W_F = -F ~\hat{n_i} \cdot \overrightarrow{\Delta r}_i,
    \label{eq:active}
\end{equation}
where, $F$ is the magnitude of the active force, representing the protrusive force due to actin polymerization \cite{Sadhu2021,Sadhu_phagocytosis} that is acting in the direction of outward normal vector of the local membrane surface (along $\hat {n_i}$) and $\overrightarrow{\Delta r}_i$ is the displacement vector of the protein complex. The ``active” forces in our simulations are implemented as external forces that act on the specific nodes of the system that contain the curved membrane proteins (with positive spontaneous curvature, red nodes in Fig.\ref{model-fig-SI}).  This is done by giving a negative energy contribution when the points on which these forces act move in the direction of the force. These forces are ``active" since they give an effective energy (work) term that is unbounded from below and thereby drive the system out-of-equilibrium. By exerting a force directed at the outwards normal we naturally describe Arp2/3-driven branching polymerization of actin, which is rather isotropic and acts as a local pressure on the membrane.

Finally, the adhesion energy due to the interaction between the vesicle and the extracellular substrate, is given by, 

\begin{equation}
    W_A = -\sum_{i'} E_{ad},
    \label{eq:adhesion}
\end{equation}

where $E_{ad}$ is the adhesion energy per node, and the sum runs over all the vertices that are adhered to the substrate \cite{samo2015,miha2019,Sadhu2021}. By `adhered vertices', we mean all such vertices, whose perpendicular distance from the adhesive surface is less than a threshold, which we chose to be equal to the length $l_{min}$, which is the unit of length in our model, and defines a minimal length allowed for a bond. Thus, the total energy of the system is given by,

\begin{equation}
    W = W_b + W_d + W_F + W_A
    \label{Wtotal}
\end{equation}

We update the vesicle with mainly two moves, (1) vertex displacement and (2) bond flip. In a vertex displacement, a vertex is randomly chosen and moved by a random length and direction, with the maximum possible distance restricted by $0.15 ~l_{min}$ (Fig. \ref{model-fig-SI}(B)). This movement provides shape fluctuations to the vesicle. In the bond flip move, a single bond is chosen,  which is a common side of two neighbouring triangles, and this bond is cut and reestablished between the other two unconnected vertices (Fig. \ref{model-fig-SI}(C)) \cite{samo2015,miha2019,Sadhu2021}. The bond flip is responsible for the lateral fluidity of the system that allows the vertices to diffuse through the membrane surface. Since our protein complexes are attached to a particular vertex, it also diffuses along with the vertex in the bond flip movement. The maximum bond length is restricted to $l_{max}=1.7 ~l_{min}$ in order to maintain self avoidance of triangulated network. We update the system using the Metropolis algorithm, where any movement that increases the energy of the system (Eq.\ref{Wtotal}) by an amount $\Delta W$ occurs with rate $\exp (- \Delta W/k_BT)$, otherwise it occurs with rate unity.

%%%%%%%%%%%%%%%%%%%%%%%%%%%%%%%%%%%%%%%%%%%%%%%%%%%%%%%%%%%%%%%%%%%%%%%%%%%%%%%%%%%%
\section{Experimental Methods}

\subsection{Migrating keratocytes on sinusoidal substrate}

\subsubsection{Cell culture}
Fish epithelial keratocytes were obtained from the scales of Central American cichlid (Hypsophrys Nicaraguensis) \cite{mohammed2019substrate,Riaz2016}. Scales were gently taken off the fish and placed in the center of a microprinted PDMS-coated glass coverslips and covered with a drop of 150 $\mu L$ of culture medium. The culture medium was composed of Leibovitz’s L-15 medium (Thermo Fisher Scientific) supplemented with 10\% fetal bovine serum (FBS, Capricorn), 1\% penicillin/streptomycin (Westburg), 14.2 mM HEPES (Sigma Aldrich) and 30\% deionized water were put on top of the scale. A glass coverslip of 22 mm in diameter was deposited on top of the scales and few drops of culture medium were added around the samples. Epithelial keratocytes were cultured in the dark at room temperature for 12 h. Keratocytes were detached from the glass coverslip by incubating with a trypsin solution (1 ml per glass slide) for 5 minutes and resuspended in 4 ml of L-15 Leibovitz complete medium. Suspended cells were then transferred to FN-coated corrugated hydrogels. All experiments were made between 2 and 8 hours after cell seeding.

\subsubsection{Fabrication of corrugated polyacrylamide hydrogels by UV-photocrosslinking}
Instead of the standard radical polymerization using catalysts such as tetramethylenediamine (TEMED) and ammonium persulfate (APS), which lead to slow polymerization times, we used an Irgacure $2959$ photoinitiator (2-Hydroxy-4'-(2-hydroxyethoxy)-2-methylpropiophenone) to polymerize hydroxypolyacrylamide (hydroxy-PAAm) hydrogels. Hydroxy-polyacrylamide (hydroxy-PAAm) hydrogels were prepared by mixing acrylamide (AAm), bis-acrylamide (bis-AAm), N-hydroxyethylacrylamide (HEA), 2-Hydroxy-4'-(2-hydroxyethoxy)-2-methylpropiophenone (Irgacure $2959$, Sigma \#410896) and deionized water. A solution composed of $2836~ \mu L$ of acrylamide (AAm, Sigma \#79-06-1) at 15\% $w/w$ in deionized water, $1943 ~\mu l$ of N,N'- methylenebisacrylamide (BisAAm, Sigma \#110-26-9) at 2\% $w/w$ in deionized water, and $1065 ~\mu L$ N-hydroxyethylacrylamide monomers at $65 ~mg/mL$ in deionized water (HEA, Sigma \#924-42-5) were mixed together in a $15 ~mL$ Eppendorf tube \cite{VERSAEVEL201433,Riaz2016,C2LC41168G} and deionized water was added to reach a final volume of $6 ~mL$. We prepared a stock solution of Irgacure $2959$ in sterile deionized water at $5 ~mg/mL$. We introduced $1 ~mL$ of the stock solution into the $6 ~mL$ of the hydrogel solution to obtain a final concentration of $0.7 ~mg/mL$. After a gentle mixing, the solution was degassed during $30~ min$ under a nitrogen flow. Glass coverslips of $22 ~mm^2$ in diameter were cleaned with $0.1 ~M$ NaOH solution during $5 ~min$ and then rinsed abundantly with deionized water during $20 ~min$ under agitation. Cleaned glass coverslips were then treated during one hour with 3-(trimethoxysilyl)propyl acrylate (Sigma \#2530-85-0) to promote a strong adhesion between the hydroxy-PAAm hydrogel and the glass coverslips and finally dried under a nitrogen flow. A volume of $40~\mu L$ of the degassed mixture was squeezed between an activated glass coverslip and a chromium optical photomask (Toppan photomask, France) and before exposition to UV illumination at $360 ~nm$ (Dymax UV light curing lamp). Chromium optical photomasks with alternating transparent stripes of $10~\mu m$ wide and black stripes of $10~\mu m$ or $20~\mu m$ wide were used to form corrugated hydrogels with wavelengths of $20~\mu m$ ($\lambda 20$), $30~\mu m$ ($\lambda 30$) and $50~\mu m$ ($\lambda 50$) and respectively \cite{luciano2021cell}. After UV exposition at 360 nm during 10 min at $10 ~mW/cm^2$ through the optical photomask, the polymerization was completed and a corrugated hydroxy-PAAm hydrogel was formed. The amplitude of $20 ~\mu m$ ($\lambda 20$) and $30 ~\mu m$ ($\lambda 30$) corrugated hydrogels was changed by adjusting the volume of the degassed polyacrylamide solution squeezed between the glass coverslip and the chromium optical photomask. Finally, hydrogels were gently removed from the photomask under water immersion, washed three times in sterile deionized water under gentle agitation and stored in sterile deionized water at $4^\circ C$.  Photocrosslinked hydroxy-PAAm hydrogels were optically transparent and did not exhibit any autofluorescence background at $470\pm20 ~nm$, $562\pm88 ~nm$ and $591\pm21 ~nm$. 

\subsubsection{Time-lapse imaging}
Time-lapse microscopy experiments were carried out on a Nikon Ti-U inverted microscope (Nikon, Japan) equipped with Differential Interference Contrast (DIC) mode. Images were taken every 3 min using a $\times 10$, $\times 20$ or $\times 40$ objective and captured with a DS-Qi2 camera (Nikon, Japan) controlled with the NIS Elements Advanced Research 4.0 software (Nikon). Tracking of the indivudal keratocytes on the corrugated hydrogels were performed with CellTracker in semi-automatic mode \cite{Piccinini2015}.

%%%%%%%%%%%%%%%%%%%%%%%%%%%%%%%%%%%%%%%%%%%%%%%%%%%%%%%%%%%%%%%%%%%%%%%%%%%%%%%%%%%

\subsection{Migration of \textit{Dictyostelium discoideum} (D. d.) cells on sinusoidal and cylindrical substrates}
\subsubsection{Cell culture}
Dictyostelium Discoedium (D.D). exists as several cell strains that need to be treated differently. An important difference between strains is the ability to use different food supplies. The cell lines used here were all axenic, meaning that they were able to feed on the culture medium (HL5 (Formedium, Norwich, England)).  To use the cells for experiments, frozen stock was thawed at room temperature and afterwards cultured in HL5 medium on Petri dishes. The doubling time of the cells was between 8 to 9 hours at the optimal growing temperature of $21-23 ^\circ C$. The cell culture was subcultured every $2-3$ days, when the cells have became confluent on in the Petri dish. The passage number was increased by one each time for a new subculture and the cells were discarded after passage 15. In the experiments we used cells harvested in their exponential growth phase. The preparation of the cells started one day before the experiment. 106 cells were pipetted into a flask with $25~ml$ HL5 medium. This flask is cultivated on a shaking table at $22^\circ C$ with $150$ rotation per minute. On the day of the experiment, 7 hours prior to the start of the experiment, the cells were centrifuged and the medium was removed. The cells are washed with phosphate buffer and afterwards centrifuged again. The remaining pellet was diluted with $20~ ml$ phosphate buffer and was positioned on the shaking table at $22^\circ C$.  Every 6 minutes a pulse of cAMP (18 Mol; Sigma-Aldrich) was delivered into the shaking culture. After six hours of starvation and the cells were chemotactically competent. AX3-ACA-Null cells were used, as the lack the aggregation stage adenylyl cyclase (ACA), i.e the cells were not able to communicate with each other. This missing functionality enabled us to investigate the effect of the complex geometry without the influence of chemotaxis. 

\subsubsection{Experiments on glass capillaries}
\begin{figure}[h!]
\centering
\includegraphics[scale=0.25]{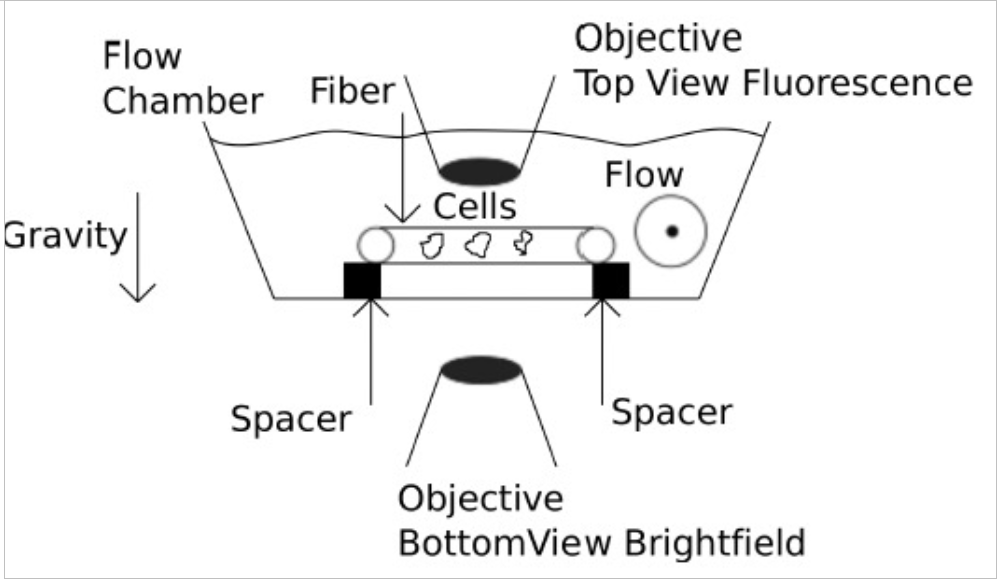}
\caption{Sketch of the curvotaxis setup. The optical fiber is placed on two glass spacers inside the perfusion chamber. The perfusion chamber is connected to a perfusion pump. Cells are added to the fiber from the top. Imaging can be done as well with inverted as with top-view optical setups}
\label{christoph-setup}
\end{figure}

The cell migration was observed on glass capillaries. To exclude any communication between the cells by signaling molecules we placed the optical fibers in a perfusion chamber (RC-27, Large Bath Chamber, Warner Instruments, Hamden, CT, USA) on a glass-spacers to allow a fluid flow around the fiber or used a microfluidic device with through flow. Fig. \ref{christoph-setup} shows the fiber setup. The cells were imaged from below with an inverted optical microscope through the number 1 cover slip. We use a peristaltic pump (RP-1 Peristaltic Pump, Mettler Toledo Inc., Columbus, Ohio USA) to create a fluid flow with a mean flow speed in the chamber of $v   = 167 \mu m/s$ . To investigate the actual velocities of the fluid flow close to the fiber, we used fluorescently labeled polymer beads Duke $36-6$ (Polystyrene Divinylbenzene (PS-DVB), Duke Scientific Corporation, California, USA), with a diameter of $33 \mu m$. The mean bead velocity close to the fiber was $v = 10 ~\mu m/s$. Hence we can be sure that the velocity in the setup did not induce shear driven migration of the D. d. cells but was still high enough to flush away any signaling molecules.  The drawbacks of this setup are the lensing effect of the fiber. See Ref. \cite{blum2015curvotaxis} for details.

\subsubsection{Experiment on sinusoidal substrate}
We use the Photonic Professional (GT) (Nanoscribe) to produce masks for sinusoidal substrates.  We chose IP-S, a highly viscous photoresist, in combination with a $25 \times$ objective (ZEISS  $25 \times$/ 0,8 DIC Imm Korr LCI Plan-NEOFLUAR) and an ITO-coated DiLL glass substrate (size $25~ mm \times 25 ~mm$; thickness $0.7 ~mm$; optical transparent; provided by NanoScribe) for the mask production. We use the 3D print as a negative and pour polydimethylsiloxane (PDMS) onto the structure where PDMS and the cross-linker (Sylgard $184$, Th. Geyer) are mixed in a mass ratio $10:1$. Finally, the PDMS, as well as an Ibidi bottom glass substrate (ibidi GmbH), are treated with a plasma cleaner provided by Harrick Plasma to clean and to oxidize the surfaces. In doing so, the PDMS sticks well to the substrate after treatment.

The optical setup and D.d. cells are pipetted onto the sinusoidal wave structures. With the help of the spinning disc confocal laser scanning microscope (sdCLSM), the cells are tracked in 3D over time to observe their amoeboid motion. The exposure time is set to $75 - 100~ ms$, the acquisition time to $351 - 376 ~ms$, the z-step is $0.8-1.5 ~\mu m$ and the number of slices is $15-25$. Stacks are recorded every $30~ s$. For details see Ref. \cite{Marcel-thesis}.
%%%%%%%%%%%%%%%%%%%%%%%%%%%%%%%%%%%%%%%%%%%%%%%%%%%%%%%%%%%%%%%%%%%%%%%%%%%%%%%%%%%%%

\subsection{Spreading and migration of \textit{Madin-Darby canine kidney} (MDCK) cells on fibers and inside tube}
\subsubsection{Microfabrication of elastomeric microtubes and microfibers}
Microtubes were fabricated inside polydimethylsiloxane (PDMS) blocks using previously described method \cite{xi2017emergent}. Briefly, a fresh mixture of silicone elastomer base and silicone elastomer curing agent (Sylgard $184$, DOWSIL$^{TM}$, $10:1$ by weight) was cured on aligned smooth copper or platinum wires (Goodfellow SARL) of different diameters. The metal wires were later pulled out leaving parallel microtubes in the PDMS block. As-fabricated PDMS blocks were then stuck to a glass-bottom petridish (FluorodishTM, Cat\#: FD$35-100$) and coated with fibronectin (Sigma-Aldrich) for cell adhesion.  

As described previously \cite{Carles2022}, PDMS cylindrical microfibers were fabricated by pulling PDMS fibers out of a pre-cured mixture of silicone elastomer base and silicone elastomer curing agent (Sylgard $184$, DOWSIL$^{TM}$, $10:1$ by weight). The mixture was mixed and left at room temperature for about 10 h before its viscosity increased to allow pulling fibers. As-fabricated microfibers were hanged in an 80$^\circ C$ oven for $1~h$ for full polymerization. The microfibers were then hanged in a glass-bottom petridish and coated with fibronectin for cell seeding.

\subsubsection{MDCK cell migration on cylindrical microfibers and microtubes}
MDCK-LifeAct-GFP (stable cell line transfected with LifeAct GFP, binding to actin filaments) cells were cultured in complete DMEM medium (Life Technologies), supplemented with 10\% fetal bovine serum and 1\% penicillin/streptomycin). Cells were cultured at $37^\circ C$ and $5\%$ CO$_2$ conditions until confluent. The cells were then collected and seeded on microfibers and microtubes at 50 million cells/mL. After $2~h$, the samples were washed carefully with clean DMEM medium to remove un-attached cells and left single, isolated cells attached on the scaffolds. 

The samples were then mounted on a confocal microscope (Zeiss, LSM $780$). To record a 3D, live-cell video, z-stacks ($1~\mu m$ per Z step) covering the whole volume of PDMS microfibers or microtubes were recorded at $10$ min/frame with either $25\times$, $40 \times$ or $63\times$ objectives. 3D time-lapse videos were recorded over a period ranging from $10$ to $24~ h$.  

For image analysis, we first converted 3D z-stack images into 2D projections as described previously \cite{xi2017emergent}. The 2D time-lapse projections were then used for PIV mapping with PIVlab (an implemented tool for MATLAB R2020) and cell tracking. To avoid bias, we flipped the 2D projected movies so that at the end of the movies, the cells' horizontal positions are on the right side in comparison to their initial position.

%%%%%%%%%%%%%%%%%%%%%%%%%%%%%%%%%%%%%%%%%%%%%%%%%%%%%%%%%%%%%%%%%%%%%%%%%%%%%%%%%%
\subsection{Migration of \textit{Dictyostelium discoideum} (D. d.) cells on micropillars}

\subsubsection{Cell culture and imaging}
The non-axenic {\it D. discoideum} strain DdB NF1 KO \cite{Bloomfield2015}, transformed with an
episomal plasmid encoding for Lifeact-GFP and PHcrac-RFP (SF108, as described in \cite{carsten2020pnas}) was used. Cells were cultivated in $10~cm$ dishes with S{\o}rensen's buffer ($14.7$ mM KH$_2$PO$_4$, $2$mM Na$_2$HPO$_4$, pH $6.0$) supplemented with $50 ~\mu M$ MgCl$_2$, $50 ~\mu M$ CaCl$_2$ and using G418 (5 $\mu g/ml$) and hygromycin ($33 \mu g/ml$) as selection markers. Klebsiella aerogenes with an $OD_{600}$ of $20$ were added to the solution in $1:10$ volume to a final $OD_{600}$ of $2.$ Before imaging, bacteria were removed by washing the cells 3 times with S{\o}rensen's buffer by centrifugation at $300 \times g$. Cells were harvested in the last washing step and transferred to the PDMS block containing the pillar structures. A glass coverslip (\#1.5, $24 \times 24~ mm$, Menzel Glaser) was used to cover the sample and prevent contamination and evaporation of the cell solution. The cells were diluted to a density that enabled imaging of single cells on the surface of the pillars. Temporal recordings were acquired at a rate of 0.2 fps using a laser scanning microscope (LSM780, Zeiss, Jena) with a 488 nm Argon laser and a $40\times$ water immersion objective.

\subsubsection{Microfabrication of pillar structures} 
A silicon wafer was coated with a $10~ \mu m$ photoresist layer (SU-8 2010, Micro Resist Technology GmbH, Germany) and patterned by direct write lithography using a maskless aligner ($\mu$MLA, Heidelberg Instruments Mikrotechnik GmbH, Germany). Polydimethylsiloxane (PDMS, Sylgard 184, Dow Corning GmbH, Germany) at a ratio of $10:1$ (base to curing agent) was spin coated on the microstructured wafer to obtain a thin ($\sim 300 \mu m$) film and cured for $2~h$ at $75^\circ C$.  A PDMS block containing the micropillars design was cut out and placed on top of a glass coverslip (\#1, $24 \times 40~ mm$, Menzel Glaser).

%%%%%%%%%%%%%%%%%%%%%%%%%%%%%%%%%%%%%%%%%%%%%%%%%%%%%%%%%%%%%%%%%%%%%%%%%%%%%%%%%
%%%%%%%%%%%%%%%%%%%%%%%%%%%%%%%%%%%%%%%%%%%%%%%%%%%%%%%%%%%%%%%%%%%%%%%%%%%%%%%%%%%%
\section{Preparation of motile vesicles on curved surfaces}
\begin{figure}[h!]
\centering
\includegraphics[scale=1.5]{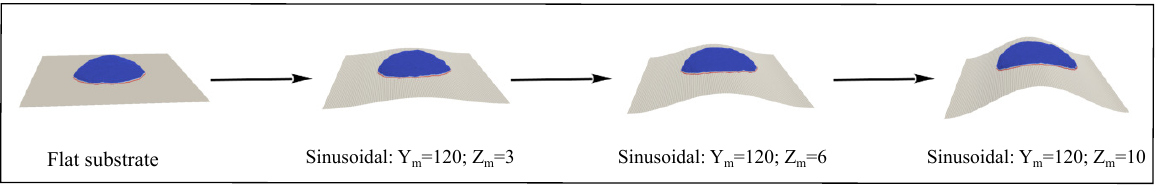}
\caption{Mechanism of preparation of vesicles on curved geometries.}
\label{adhere-slowly}
\end{figure}

In our simulation, we often obtain a migrating vesicle on curved surfaces (cylinder or sinusoidal) by spontaneous polarization of proteins when started from a spherical-like vesicle. In this case, however, the direction of polarization is random and set by the minimum of energy. So, we could not allow the vesicle to migrate in a direction of our choice. In order to start with a vesicle that migrate in a direction of our choice, we thus use a migrating vesicle already generated on a flat substrate, and place it on a curved surface and allow it to adjust on the curved substrate.

The above mentioned method enable us to study the migration of vesicle of a direction of our choice, but often, the cluster of the migrating vesicle breaks into parts, since at the time of replacing the substrate (flat into curved) some of the vertex containing proteins detaches from the substrate. In order to resolve this issue, we use another method, where we slowly change the curvature of the substrate and allow the vesicle to adjust on the substrate without detaching the vertices from the substrate (Fig. \ref{adhere-slowly}). The process of adhering on the substrate is much faster than the reorientation of the vesicle, such that the vesicle does not rotate much while adhering to the curved substrate.
%%%%%%%%%%%%%%%%%%%%%%%%%%%%%%%%%%%%%%%%%%%%%%%%%%%%%%%%%%%%%%%%%%%%%%%%%%%%%%%%%%%%%
\section{More trajectories for the migration on sinusoidal substrate with large wavelength}
\begin{figure}[h!]
\centering
\includegraphics[scale=1.5]{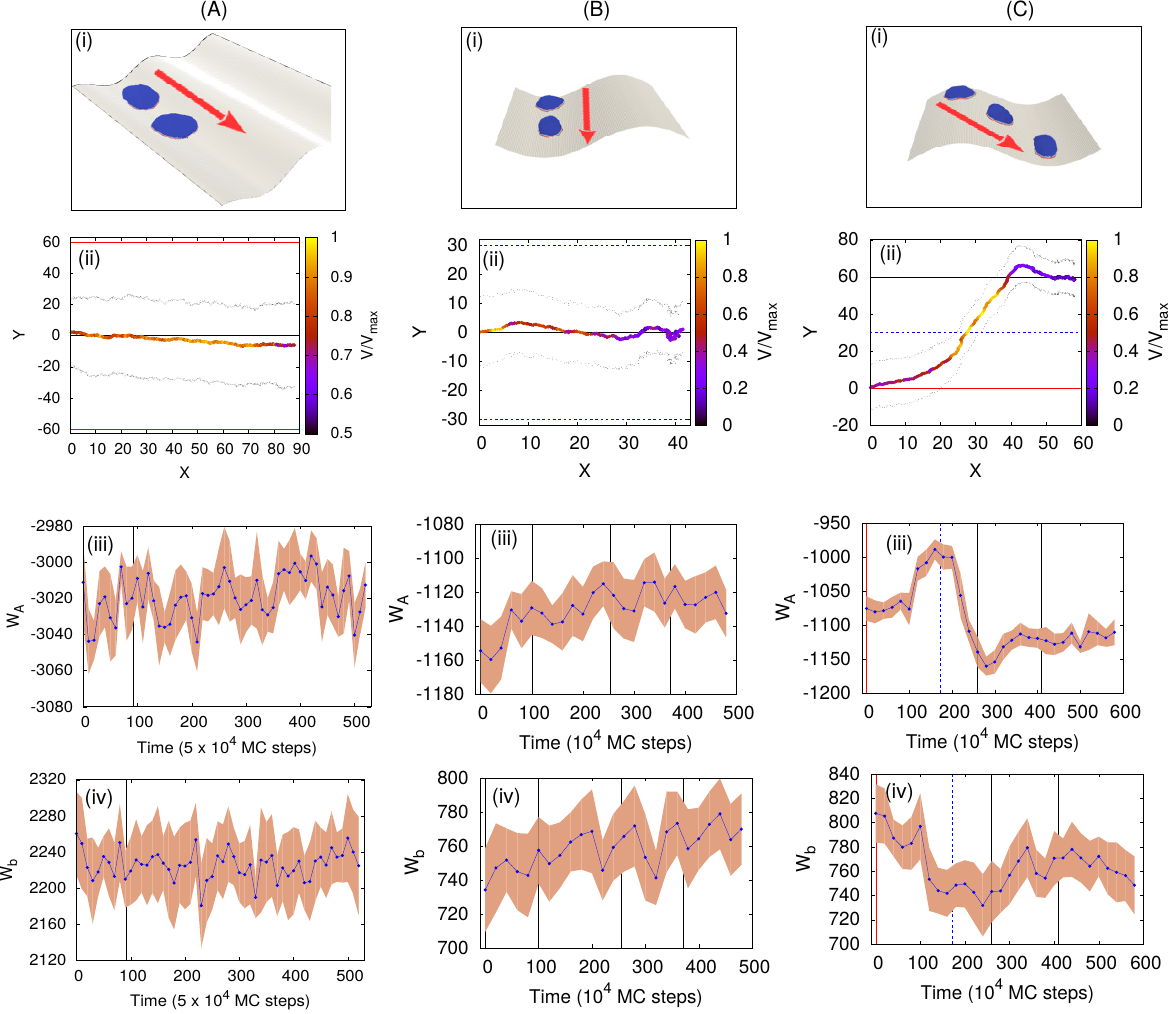}
\caption{More trajectories for motile vesicle moving on a sinusoidal substrate with large wavelength. We use $z_m=10~l_{min};~y_m=120~l_{min}$. (A) A large vesicle starting from the minimum of the sinusoidal substrate continues its initial direction of migration. (B) A small vesicle starting from the minimum of the sinusoidal substrate continues to the minimum of the substrate. (C) Small vesicle starting from the maximum of a sinusoidal  substrate shifts to the minimum. In (i), we show the configurations, in (ii), we show the trajectories with the speed of the vesicle in color code, in (iii), we show the adhesion energy and in (iv) , we show the bending energy. For large vesicle (A), we use $N=3127$, $E_{ad}=2.0$ $F=1.0 k_BT/l_{min}$ and $\rho=2.4 \%$. For small vesicle (B-C), we use $N=607$, $E_{ad}=3.0~k_BT$, $F=2.0 k_BT/l_{min}$ and $\rho=4.9\%$.}
\label{ym_by_Rv_large}
\end{figure}

Here, we show few more trajectories of vesicle migrating on sinusoidal substrate with large wavelength. In Fig. \ref{ym_by_Rv_large}(A), we show a large vesicle starting from the minimum of the sinusoidal substrate maintains its direction of motion, and finally looses its motility property after long time. Similar behaviour is observed for a small vesicle starting from the minimum as shown in Fig. \ref{ym_by_Rv_large}(B). A small vesicle, when starting from the ridge quickly slides down and then moves along the axis maintaining its direction of motion (Fig. \ref{ym_by_Rv_large}(C)).

The corresponding trajectories are shown in second panel (Fig. \ref{ym_by_Rv_large} (ii)). We also show the adhesion and bending energise for each case in Fig. \ref{ym_by_Rv_large}(iii-iv), that shows similar behaviour as for the other cases.
%%%%%%%%%%%%%%%%%%%%%%%%%%%%%%%%%%%%%%%%%%%%%%%%%%%%%%%%%%%%%%%%%%%%%%%%%%%%%%%%%%%%%
\section{Variation of protein-protein binding energy for the migration on sinusoidal substrate with large wavelength}
Here, we show the variation of protein-protein binding energy with time when  a large vesicle migrates on a sinusoidal substrate with $z_m=10~l_{min}, ~y_m=120~l_{min}$ (Fig. 2(C-D), main paper). We note that the energy is minimum when the vesicle is on the maximum of the substrate. This indicates that the proteins form much more strong cluster when the vesicle is on the maximum of the sinusoidal substrate. However, the variation in the protein-protein binding energy is much smaller in comparison with the other energies. 
\begin{figure}[h!]
\centering
\includegraphics[scale=0.7]{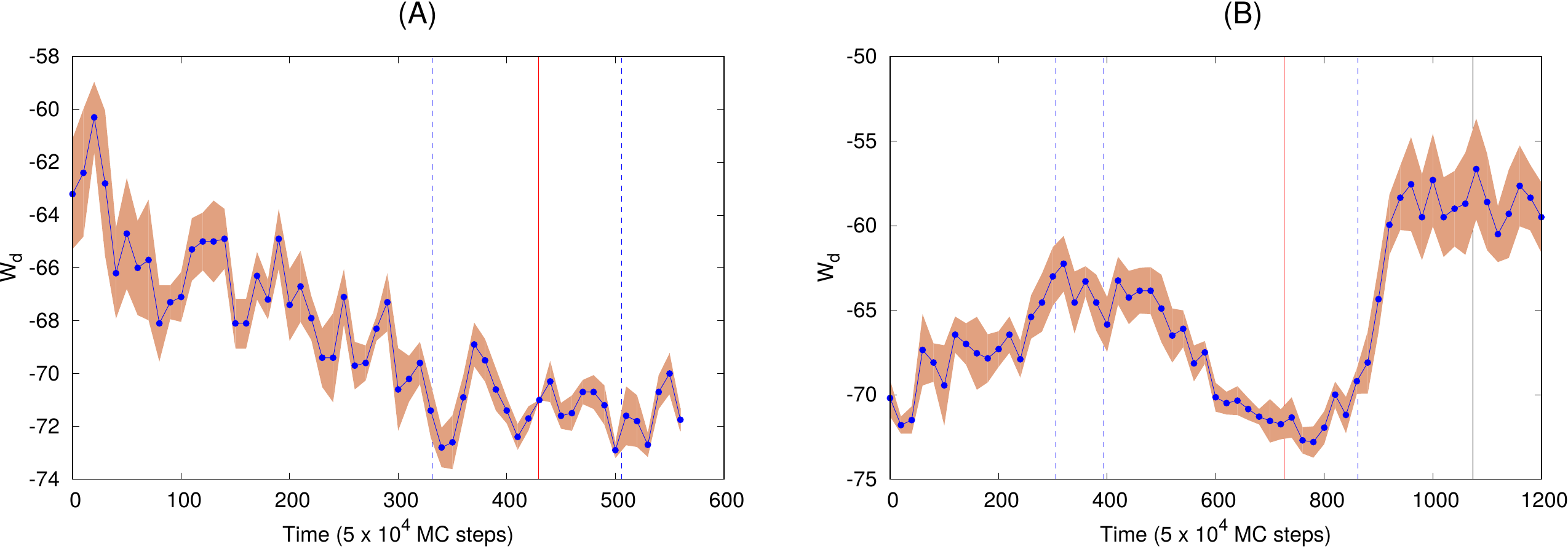}
\caption{Variation of protein-protein binding energy when a large vesicle migrates on a sinusoidal substrate with $z_m=10~l_{min}, ~y_m=120~l_{min}$. For (A), we use $F=2.0~k_BT/l_{min}$ and for (B), we use $F=1.0~k_BT/l_{min}$. Other parameters are, $N=3127$, $E_{ad}=2.0~k_BT$, $\rho=2.4~\%$}
\label{mean-custer}
\end{figure}
%%%%%%%%%%%%%%%%%%%%%%%%%%%%%%%%%%%%%%%%%%%%%%%%%%%%%%%%%%%%%%%%%%%%%%%%%%%%%%%%%%%
\section{Periodicity in the speed for the migration on sinusoidal substrate with small wavelength}
\begin{figure}[h!]
\centering
\includegraphics[scale=1.1]{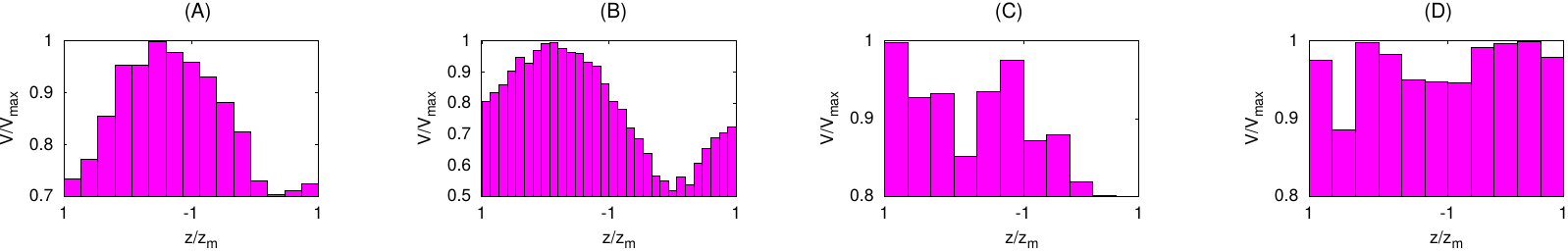}
\caption{Periodicity in the speed for the migration on sinusoidal substrate with small wavelength (Fig. 3, main text). We plot the distribution of speed in a full period of sinusoidal variation for the case of Fig. 3, main text, for each of the four cases respectively. Here, $z/z_m=1$ represents the ridges (maximum) while $z/z_m=-1$ represents the grooves (minimum) of the sinusoidal substrate. For simulation results, we use $N=607$, $E_{ad}=3.0~k_BT$, $F=4.0 k_BT/l_{min}$ and $\rho=4.9\%$.}
\label{sinusoidal-periodicity}
\end{figure}

In Fig. \ref{sinusoidal-periodicity}, we show the distribution of speed in a full period of sinusoidal variation for the case of Fig.3 from the main text, for each of the four cases. Fig. \ref{sinusoidal-periodicity}(A-B) shows nice periodicity for the simulation results, but Fig. \ref{sinusoidal-periodicity}(C-D), does not show nice periodic variation from the migration of karatocytes, as the errors are larger in the case of experiment. 

%%%%%%%%%%%%%%%%%%%%%%%%%%%%%%%%%%%%%%%%%%%%%%%%%%%%%%%%%%%%%%%%%%%%%%%%%%%%%%%%%%%%%
\section{More trajectories for the migration on sinusoidal substrate with small wavelength}
\begin{figure}[h!]
\centering
\includegraphics[scale=1.05]{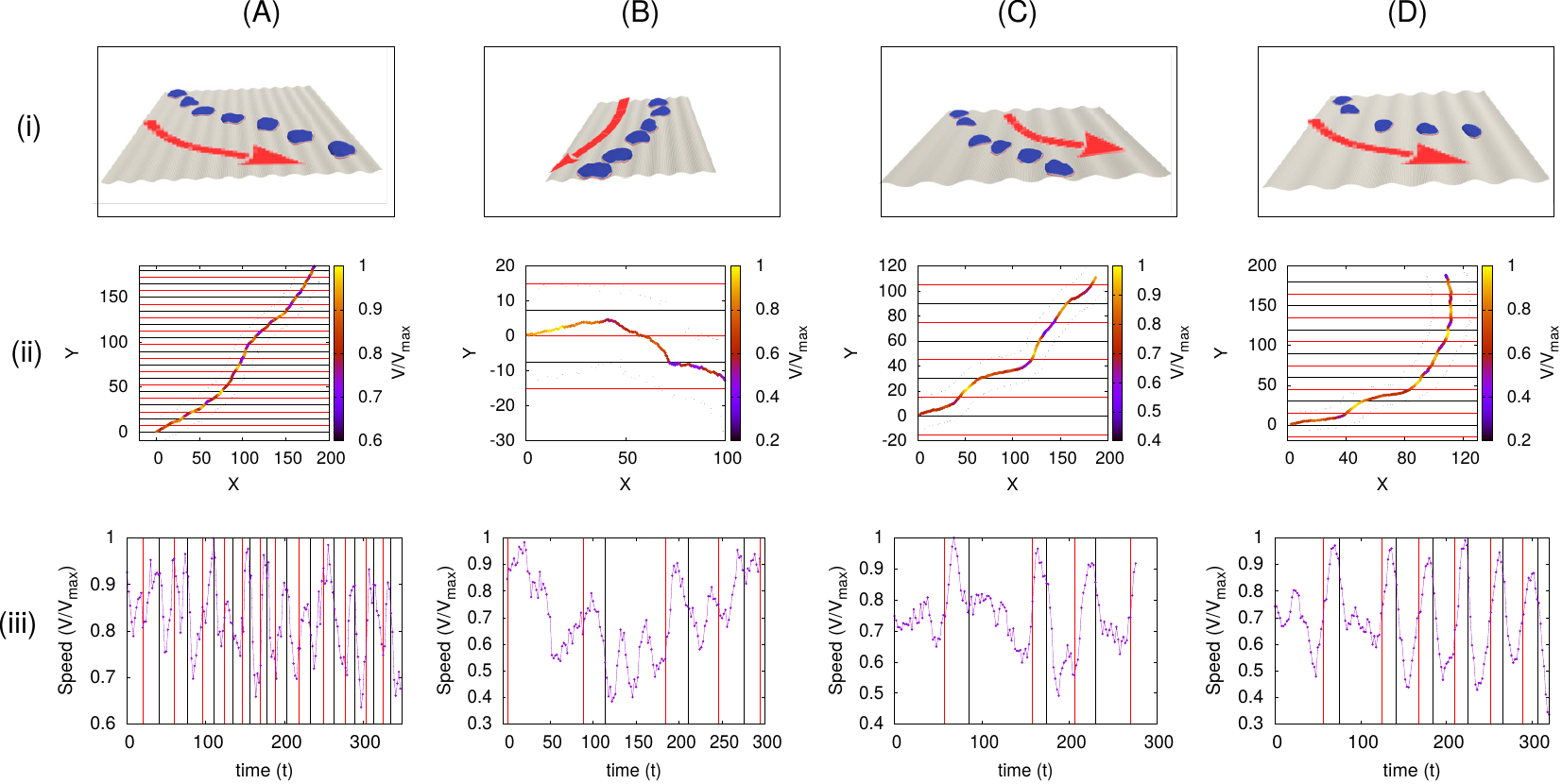}
\caption{More trajectories for motile vesicle moving on a sinusoidal substrate with small wavelength. (A) A vesicle started from  the minimum of the substrate shows zig-zag like motion with oscillating speed along the trajectory. (B) A vesicle started from the maximum of the sinusoidal substrate migrates with oscillating speed. (C-D) Two different simulations of vesicle starting from the minimum. For (A-B) we use  $z_m=1~l_{min};~y_m=15~l_{min}$ for the sinusoidal substrate, and for (C-D), we use  $z_m=2~l_{min};~y_m=30~l_{min}$. Other parameters are $N=607$, $E_{ad}=3.0~k_BT$, $F=4.0 k_BT/l_{min}$ and $\rho=4.9\%$. }
\label{ym_by_Rv_small}
\end{figure}

Here, we show few more trajectories for the sinusoidal substrate with small wavelength using the small vesicle only. In \ref{ym_by_Rv_small}(A-C), the vesicle does not maintain any particular direction and shows zig-zag like migration. In \ref{ym_by_Rv_small}(D) the vesicle finally migrates in a direction orthogonal to the sinusoidal substrate.

In second panel (\ref{ym_by_Rv_small}(ii)), we show the trajectories for each cases, and in the third panel (\ref{ym_by_Rv_small}(iii)) we show the speed showing oscillatory behaviour similar to the previous cases.

%%%%%%%%%%%%%%%%%%%%%%%%%%%%%%%%%%%%%%%%%%%%%%%%%%%%%%%%%%%%%%%%%%%%%%%%%%%%%%%%%%%%%
\section{More trajectories of Karatocytes migrating on sinusoidal substrates}
\begin{figure}[h!]
\centering
\includegraphics[scale=0.75]{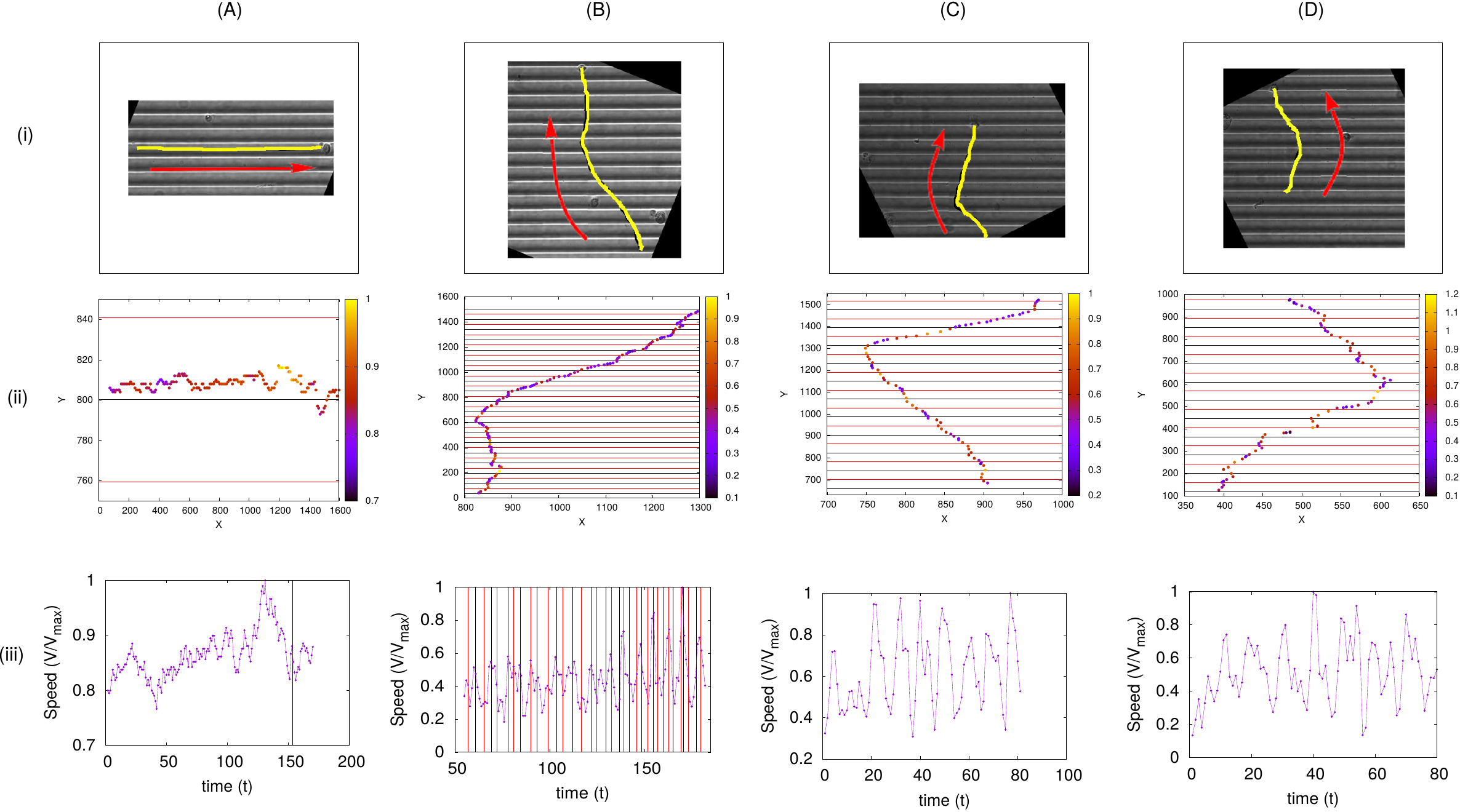}
\caption{Here, we show few more trajectories for the migration of \textit{karatocytes} on sinusoidal substrates. (A) The cell seems to move along the minimum without tending to change its direction of migration. (B-D) The cell migrates almost orthogonal to the sinusoidal substrate throughout its trajectories. Here, in (i) we show the snapshots, in (ii) we show the trajectories with the speed in the colour code, and in (iii), we show the speed of the cell with time, showing oscillatory behaviour. }
\label{karatocyte}
\end{figure}

Here, we show the few more trajectories of the migration of karatocytes on sinusoidal substrate in  Fig. \ref{karatocyte}. In panel (i), we show the configurations, in second panel we show the corresponding trajectories and finally in the last panel of Fig. \ref{karatocyte}, we show the speed of the migrating cells with time, showing oscillatory behaviour. 

%%%%%%%%%%%%%%%%%%%%%%%%%%%%%%%%%%%%%%%%%%%%%%%%%%%%%%%%%%%%%%%%%%%%%%%%%%%%%%%%%%%%
\section{Vesicle migrating on cylindrical fiber: mapping from `time' to `migration angle'}
\begin{figure}[h!]
\centering
\includegraphics[scale=0.7]{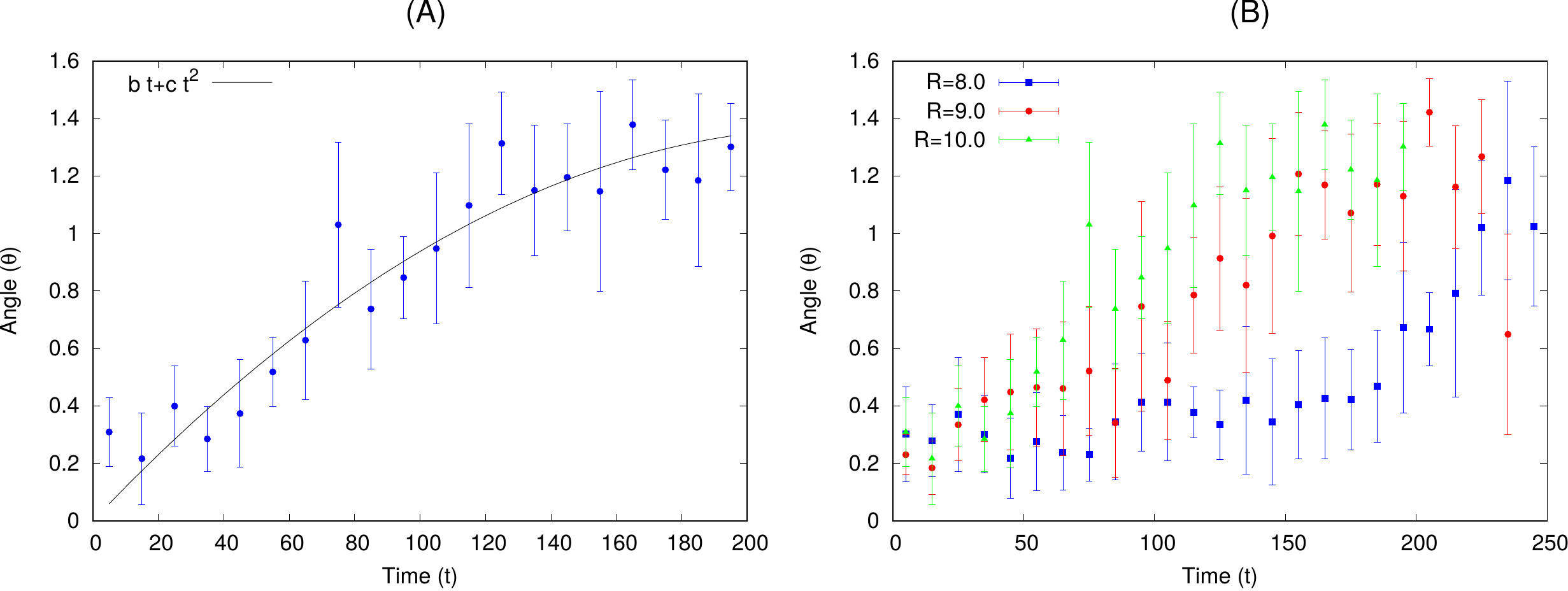}
\caption{Migration angle with time for the case of vesicle migrating on a cylindrical fiber (Fig.4A). We define this quantity as the angle between the displacement vector of the vesicle in a small time $\Delta t$ and the cylindrical axis. (A) Variation of migration angle ($\theta$) with time for $R=10~l_{min}$. We fit this data with a monotonic quadratic function of the type $bt+c t^2$, where, $b=1.20085 \times 10^{-2}$ and $c=-2.63639 \times 10^{-5}$. (B) Migration ange ($\theta$) for different values of $R$. The parameters are same as Fig. 4(A) of the main text.}
\label{fitting-t2angle}
\end{figure}

For the case of vesicle migrating on cylindrical fiber, we measure the quantities as a function of time, however, in MC simulations, the time is not a real quantity, rather it represents the MC time. So, we map time to a real quantity, i.e., the migration angle, which is defined as the angle between the displacement vector of the vesicle in a small time $\Delta t$ and the cylindrical axis. This quantity, initially starts from zero saturates to $\pi/2$ for larger times. We show this angle with time in Fig. \ref{fitting-t2angle}.

Since the measurement of angle ($\theta$) is associated with some statistical errors, we fit this data with a close quadratic function of the type $b t+ c t^2$, that monotonically increases with time. This fitting allows us to describe different quantities as a function of migration angle, rather that unrealistic MC time.

%%%%%%%%%%%%%%%%%%%%%%%%%%%%%%%%%%%%%%%%%%%%%%%%%%%%%%%%%%%%%%%%%%%%%%%%%%%%%%%%%%%%
\section{Vesicle migrating on cylindrical fiber: various energies as a function of time}
\begin{figure}[h!]
\centering
\includegraphics[scale=0.45]{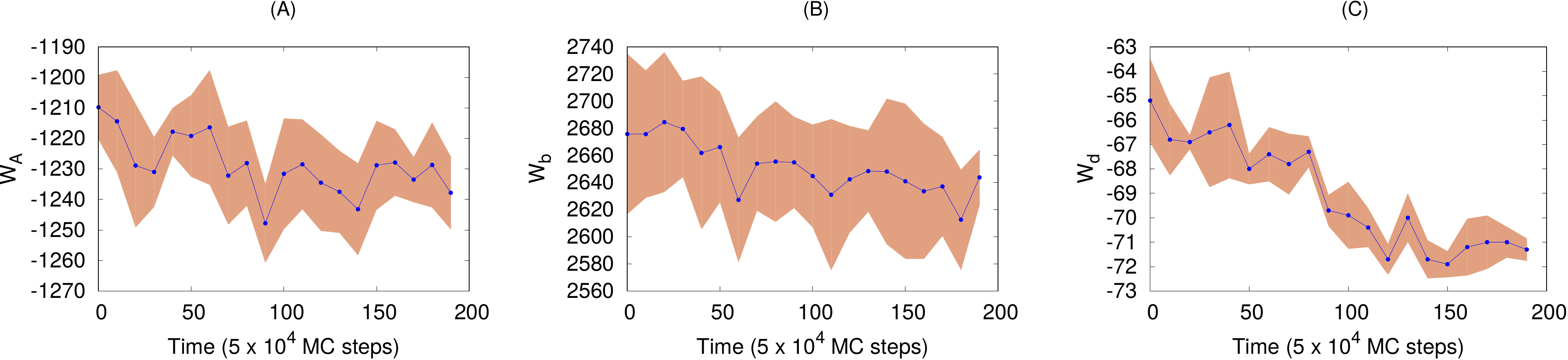}
\caption{Different energies with time for a vesicle migrating on cylindrical fiber (Fig.4A). (A) Adhesion energy with time. (B) Bending energy with time. (C) Protein-protein binding energy with time. The parameters are same as Fig. 4(A) of the main text.}
\label{t-vs-energy-fiber}
\end{figure}

In the main text (Fig. 4(B-D)), we show the variation of different energies of the vesicle as a function of its migration angle, when migrating on cylindrical fiber. Here, we present the same result as a function of MC time in Fig. \ref{t-vs-energy-fiber}.

%%%%%%%%%%%%%%%%%%%%%%%%%%%%%%%%%%%%%%%%%%%%%%%%%%%%%%%%%%%%%%%%%%%%%%%%%%%%%%%%%%%%%
\section{Energy of axial and circumferential orientation with fiber radius}
\begin{figure}[h!]
\centering
\includegraphics[scale=0.7]{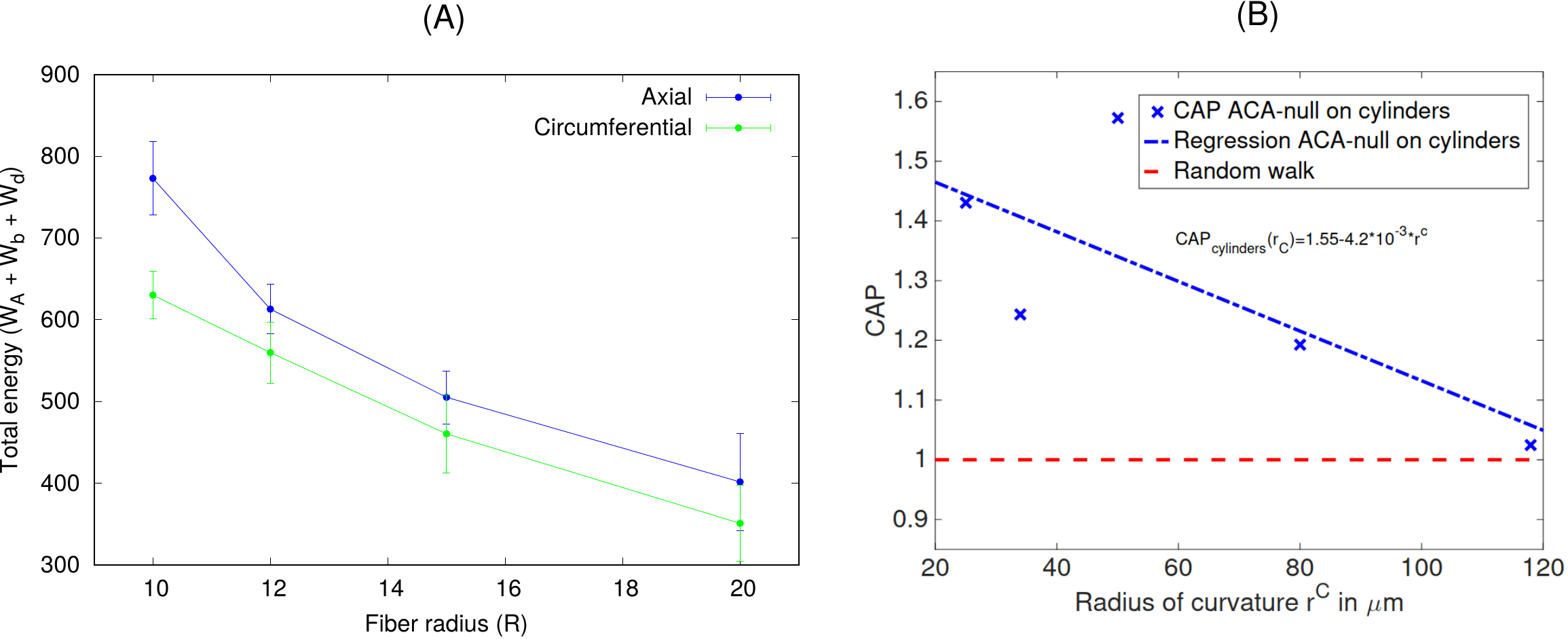}
\caption{Quantification of the anisotropy of the migration of cells/vesicles on fibers of different radii. (A) The total energy of the vesicle as a function of radius of the fiber, comparing two cases: (1) When the vesicle is migrating along the axial direction of the fiber, and (2) When the vesicle is migrating along the circumferential direction. (B) The Curvotactic Anisotropy Parameter (CAP) as a function of radius of curvature of the fiber. For simulation results, the parameters are same as in Fig. 4A of the main text.}
\label{cylinder-diff-R}
\end{figure}

Here, we show the total energy of the vesicle when it is in the axial orientation verses when it is in the circumferential orientation Fig.\ref{cylinder-diff-R}(A). We note that the energy decreases with $R$, as the curvature of the cylinder decreases with $R$. Also, the difference between the energies in the axial and circumferential orientation also decreases with $R$. 

In order to identify the anisotropy in the experimental data for D.d. cells migrating on fibers, we define the quantity called Curvotactic Anisotropy Parameter (CAP) as the absolute value of the velocity $|V_{||}|$ in the curved direction divided by the absolute value of the velocity in the perpendicular direction $|V_\bot|$,

$$CAP = \frac{<|V_{||}|>}{<|V_\bot|>}$$

In Fig. \ref{cylinder-diff-R}(B), we show this quantity for fibers of different radius of curvature. The CAP is found to be lower for larger radius of curvature, in agreement with the simulations that predict a smaller bias for circumferential orientation as the radius of the fiber increases (Fig. \ref{cylinder-diff-R}A).

%%%%%%%%%%%%%%%%%%%%%%%%%%%%%%%%%%%%%%%%%%%%%%%%%%%%%%%%%%%%%%%%%%%%%%%%%%%%%%%%%%%%%
\section{MDCK cells on fiber and inside tube}
\begin{figure}[h!]
\centering
\includegraphics[scale=1.0]{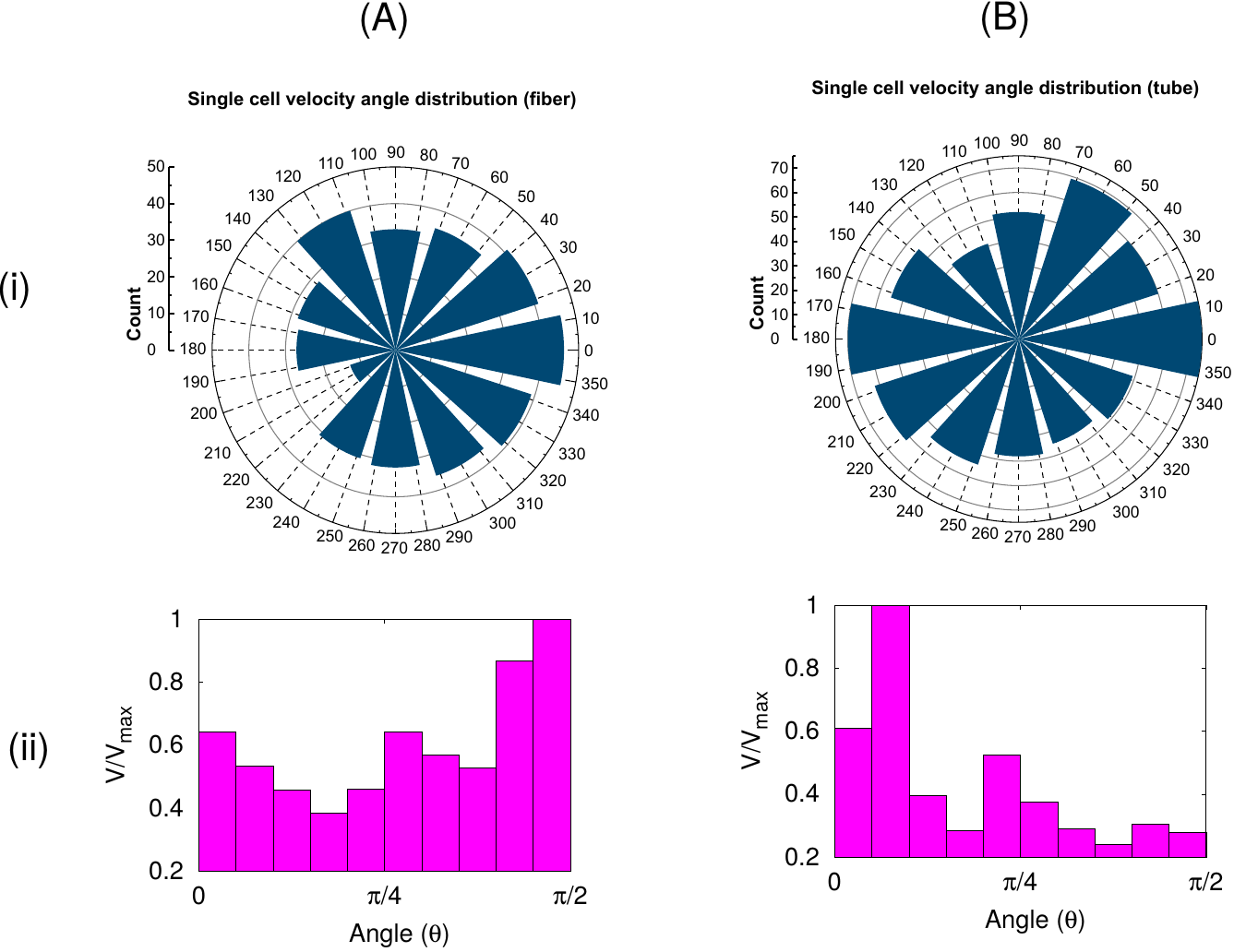}
\caption{Angular speed distribution for MDCK cells migrating on (A) Fibers of $50$ to $70 \mu m$ in diameter, and (B) Tubes of $67$ to $75 \mu m$ in diameter. The direction $0^\circ-180^\circ$ represents the circumferential direction and the direction along $90^\circ-270^\circ$ is the axial direction. }
\label{MDCK}
\end{figure}

Here, we show the distribution of velocity of the migrating MDCK cells on the fiber (Fig. \ref{MDCK}(A)) and inside the tube (Fig. \ref{MDCK}(B)).

%%%%%%%%%%%%%%%%%%%%%%%%%%%%%%%%%%%%%%%%%%%%%%%%%%%%%%%%%%%%%%%%%%%%%%%%%%%%%%%%%%%%%
\section{Vesicle migrating on cylinder of elliptical cross-section with different ratio of $R_x/R_y$}
\begin{figure}[h!]
\centering
\includegraphics[scale=0.13]{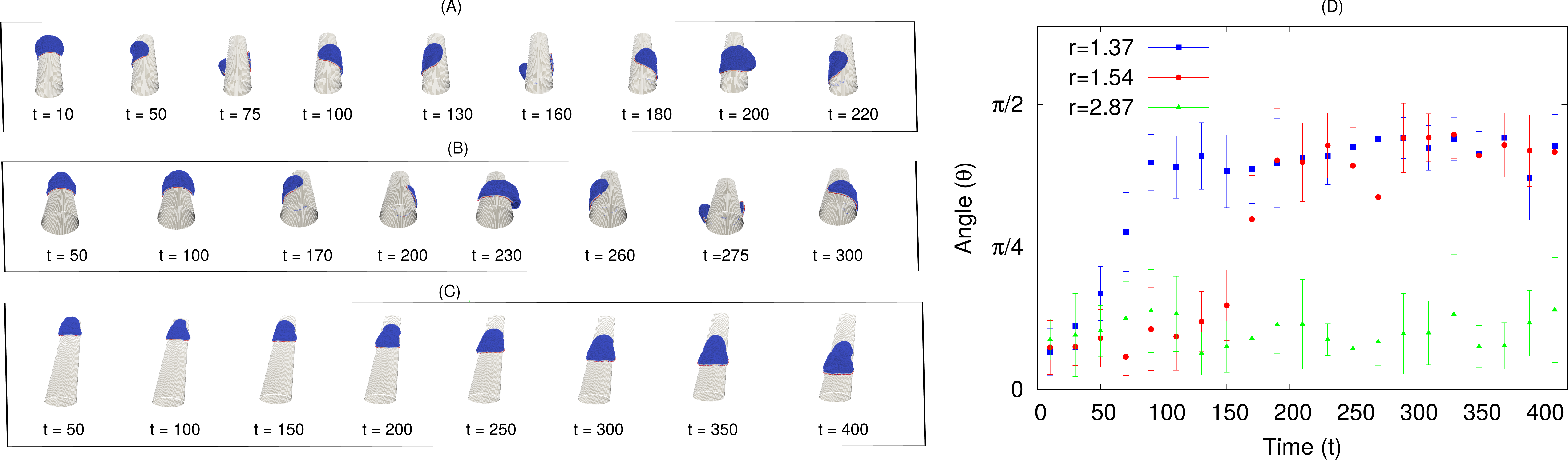}
\caption{Vesicle migrating on elliptical fiber of different aspect ratio $r=R_x/R_y$ with fixed circumference $2 \pi R$, with $R=10~l_{min}$. (A) Configuration with time for aspect ratio $r=1.37$. (B) Configuration with time for aspect ratio $r=1.54$. (C) Configuration with time for aspect ratio $r=2.87$. (D) Angle of migration $\theta$ with time for a vesicle migrating on elliptical fiber with different values of $r$. Other parameters are $E_{ad}=1.5~k_BT$, $F=2.0~k_BT/l_{min}$,  and $\rho=2.4~\%$.}
\label{elliptical-diff-r}
\end{figure}

 We note that the migrating vesicle, when exposed to a fiber of elliptical cross-section migrates and reorients in the circumferential direction when the aspect ratio $r=R_x/R_y$ is close to unity (Fig. \ref{elliptical-diff-r}A). As the aspect ratio $r$ increases, the reorientation time also increases (Fig. \ref{elliptical-diff-r}B). For a very large value of $r$, the vesicle only migrates along the axis and never rotates circumferentially (Fig. \ref{elliptical-diff-r}C). We also show the angle of migration $\theta$ with time for three different values of $r$ in Fig. \ref{elliptical-diff-r}D. 

%%%%%%%%%%%%%%%%%%%%%%%%%%%%%%%%%%%%%%%%%%%%%%%%%%%%%%%%%%%%%%%%%%%%%%%%%%%%%%%%%%%%%
%\section{Vesicle speed with different cylindrical radius}
%\begin{figure}[h!]
%\centering
%\includegraphics[scale=0.45]{angle_with_time_diff_R.eps}
%\includegraphics[scale=0.45]{coiling_speed_with_R.eps}
%\includegraphics[scale=0.45]{t_vs_alignment_crescent_along_x_R_10.eps}
%\caption{(a) Angle between the direction of motion and the axis of cylinder, for different values of $R$. (b) Coiling speed of a crescent-shaped vesicle (scaled by the speed at a flat substrate) with cylindrical radius $R$. (c) The ratio of the component of force along axial direction to circumferential direction. The parameter values are: $E_{ad}=1.0$, $F=2.0$ and $\rho=2.4 \%$.}
%\label{rotation_speed}
%\end{figure}
%%%%%%%%%%%%%%%%%%%%%%%%%%%%%%%%%%%%%%%%%%%%%%%%%%%%%%%%%%%%%%%%%%%%%%%%%%%%%%%%%%%%%%%%%%%
\section{Spreading and migration of vesicle inside a cylindrical tube}
\begin{figure}[h!]
\centering
\includegraphics[scale=1.4]{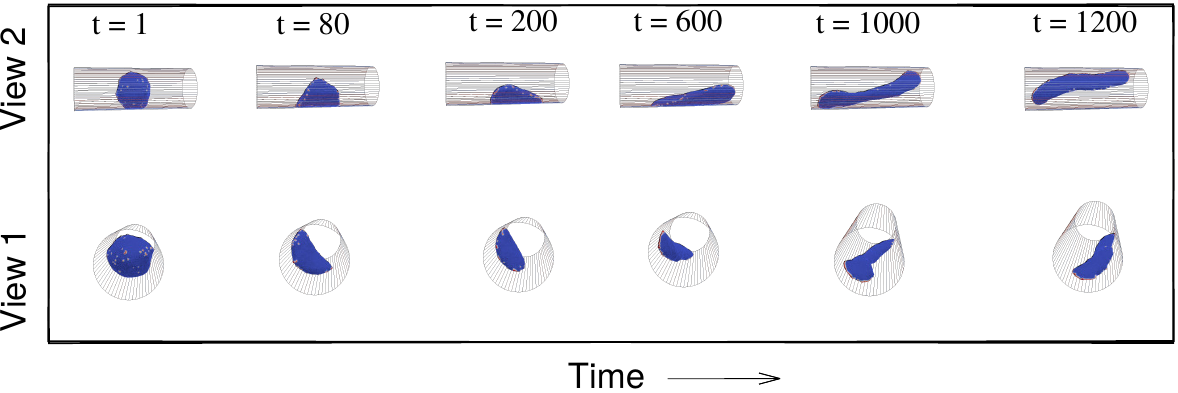}
\caption{Vesicle inside a cylinder: two arc formation. We show the configurations of a vesicle inside a cylinder, starting from a quasi-spherical vesicle. Here, we use $R=23$. Other parameter values are same as Fig. 6A of the main text.}
\label{vesicle_inside_cylinder_spreading}
\end{figure}

Here, we show the results for the (non-motile) vesicle spreading inside a cylindrical tube (Fig. \ref{vesicle_inside_cylinder_spreading}). The vesicle seems to elongate in the axial direction without any preference to orient circumferentially.

\begin{figure}[h!]
\centering
\includegraphics[scale=1.35]{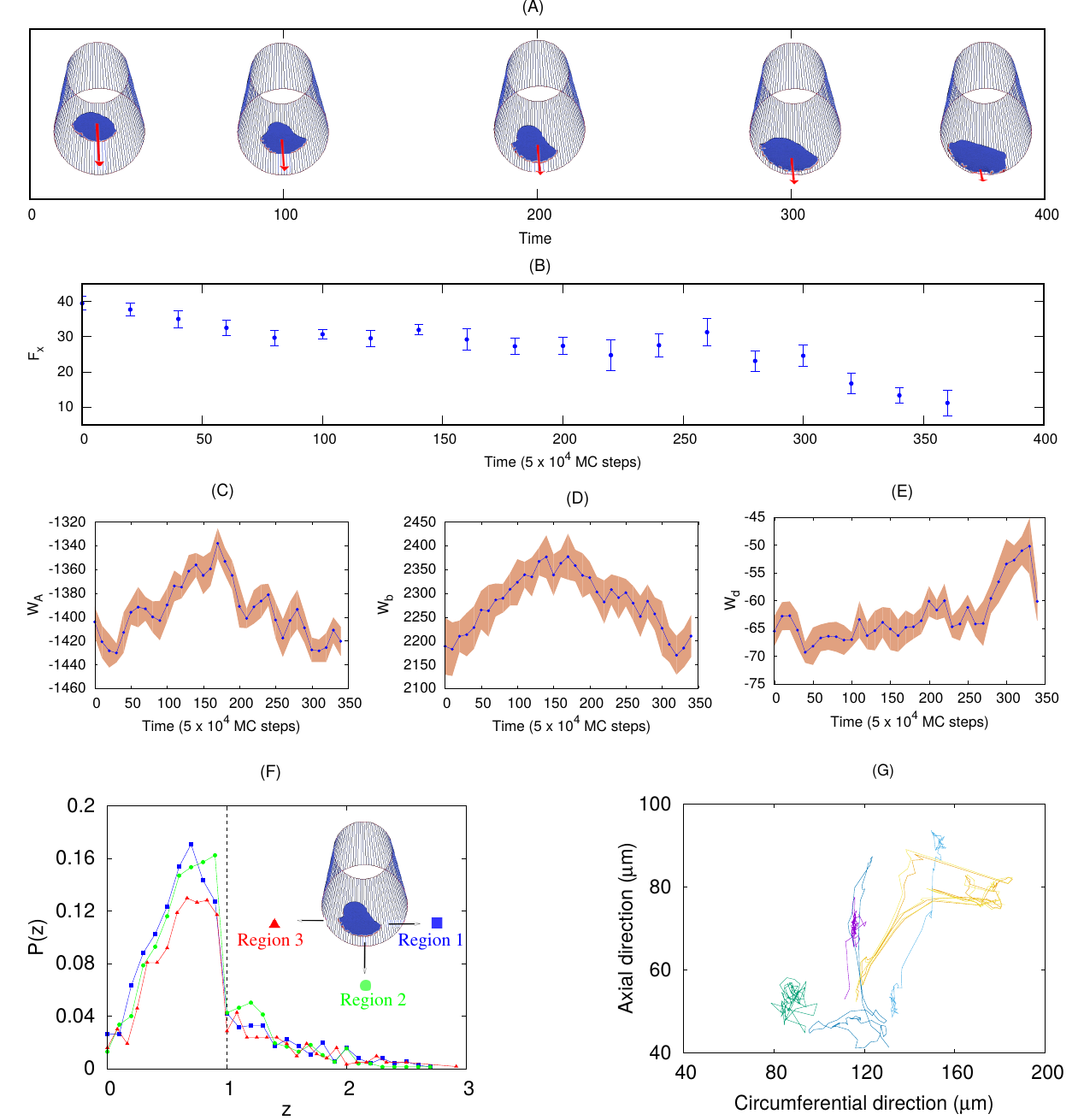}
\caption{Vesicle migrating inside a cylindrical tube. (A) Configuration of vesicle migrating inside tube, initiated in the axial direction. (B) Magnitude of active force along the axis of cylinder  ($F_x$) with time. (C) The adhesion energy of the vesicle with time. (D) The bending energy of the vesicle with time. (E) The binding energy between proteins with time. (F) Probability distribution of a protein at $z$ distance above the cylindrical substrate. The left side of the vertical dashed line (at $z=1$) represents adhered proteins. Here, we use $R=35~l_{min}$, $E_{ad}=1.0$, $F=2.0$, and $\rho=2.4 \%$. (G) Trajectories of the MDCK cells migrating inside tubes of $67-75~\mu m$ in diameter.}
\label{vesicle_inside_cylinder}
\end{figure}

Next, we study the migration of cells inside a cylindrical tube, a uniformly concave surface. We start with our vesicle, that shows no tendency to rotate circumferentially when initially aligned to migrate along the axial direction (Fig.\ref{vesicle_inside_cylinder}A, Movie-S29). Over time, the vesicle is found to lose its motility, and the leading edge protein cluster breaks into several parts, which leads to a decrease in the total active force that propels the vesicle (Fig.\ref{vesicle_inside_cylinder}B). This is similar to the behavior inside the grooves of the sinusoidal surface (Fig.2A,B, main text). We chose here a tube radius such that the circumference of the tube is much larger than the vesicle's diameter. In this regime we can explore the cell migration on the surface, avoiding ``plugging" of the tube by the vesicle when the tube radius is smaller than the cell radius \cite{xi2017emergent}.

The different energy terms of the vesicle do not show any systematic variation during its migration in the tube (Fig.\ref{vesicle_inside_cylinder}C-E). In Fig.\ref{vesicle_inside_cylinder}F we show that all the proteins along the cell edge are well adhered to the substrate, which is why the leading edge can easily break up and form clusters along any direction. The leading-edge that was initially oriented axially, will tend to break into two arcs that point side-ways. This destabilizes the polarized leading-edge aggregate, and the cell migration slows down, sometimes to a halt (forming a two-arc non-motile phenotype). As a result, we find that the vesicle in the tube ends up either as slowly migrating in the axial direction (on average, Fig.\ref{vesicle_inside_cylinder}A) or a two-arc (non-motile) vesicle that can be axial or ``bridge" orthogonally to the axis (Fig. \ref{vesicle_inside_cylinder_spreading}, Movie-S30). 

Comparing to experimental observations of cells moving inside tubes \cite{van2022myofibroblast}, it was indeed found that cells tend to migrate along the tube axis, as we obtain (Fig.\ref{vesicle_inside_cylinder}A). However, this tendency is strongly cell-type dependent, with some cells becoming non-motile inside tubes, forming adhesion "bridges" that can be orthogonal to the tube axis \cite{Werner2019}. This observation agrees with our finding that the motility inside the tube is strongly inhibited, with the cells tending to lose their polarization (Fig.\ref{vesicle_inside_cylinder}A,B). In \cite{xi2017emergent}, it was indeed observed that the motility along the tube axis decreases as the tube radius decreases, as cells migrate less persistently, and their leading-edge lamellipodia becomes less persistent and less stable. In addition, the overall orientation of actin filaments for cells inside tubes was axial, in qualitative agreement with our model's results.

Fig.\ref{vesicle_inside_cylinder}G, we show the trajectories for the migration of MDCK cells on the inside of a tube. The cells were found to be weakly motile, with the most persistent motility periods aligned with the tube axis (Movie-S31), in agreement with the model predictions.

Another recent study \cite{Yu2022} on two cell types (endothelial and epithelial cells), found similar axial alignment of the cells shape and their migration inside tubes. However, the role of stress-fibers, which we do not include in our model, was suggested to have a major role for these cells.

\begin{figure}[h!]
\centering
\includegraphics[scale=1]{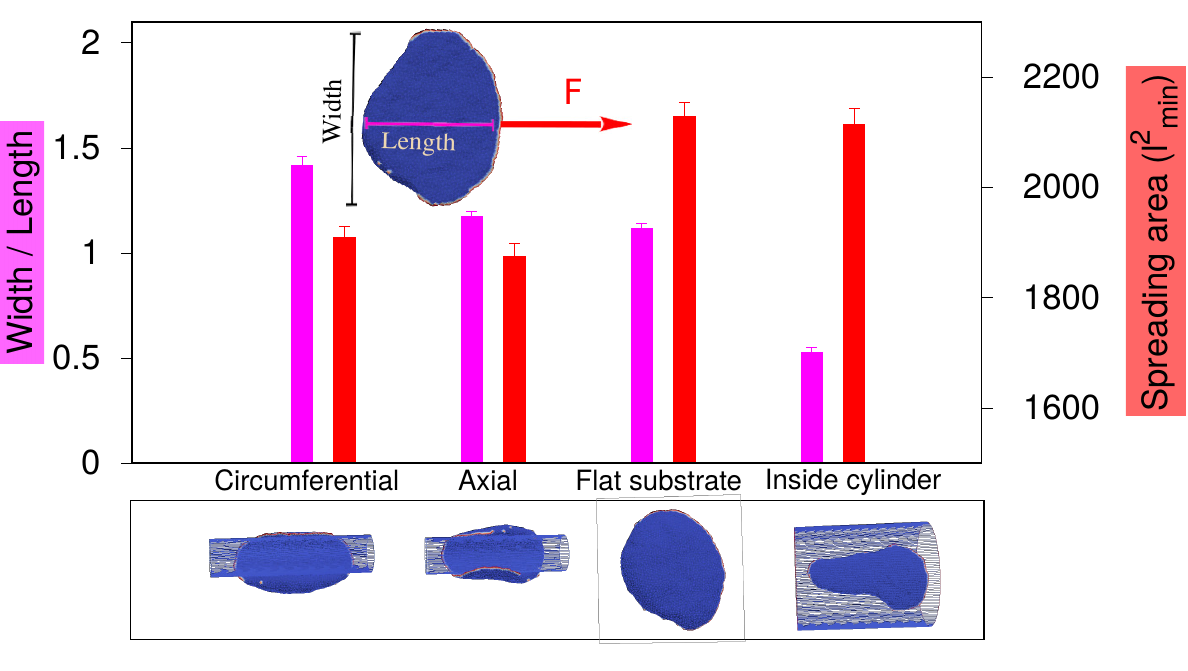}
\caption{Ratio of the length (along $F$) and width (orthogonal to $F$) of the vesicle and the adhered area, for four different cases when it is moving (1) along circumferential direction on a cylinder, (2) along axial direction on a cylinder, (3) on a flat substrate and (4) inside of a cylindrical tube. The magenta color is for the aspect ratio and the red color showing the adhered area. Inset shows a crescent shape on a flat substrate, defining the length and the width of the vesicle. For fiber, we use $R=10~l_{min}$, for tube, we use $R=21~l_{min}$. Other parameters are $F=2.0$, $E_{ad}=1.0$ and $\rho=2.4 \%$.}
\label{aspect-ratio}
\end{figure}

In Fig. \ref{aspect-ratio} we summarize the steady-state shapes for the simulated motile vesicles on the different curved substrates. We note that the adhered area is maximal for a flat substrate, but is not very different when the vesicle is inside the tube, where it is also well adhered. When on the fiber, moving in the axial direction, the adhered area is the smallest, and it is slightly larger when moving in the circumferential direction. The aspect ratio shows that the active force is able to stretch the vesicle sideways along the axis of the tube, when it is oriented circumferentially. The highly elongated shape of the aligned cell moving in the tube, compared to the flat substrate, fits well with the observed elongation of the cells when migrating inside the grooves of the sinusoidal substrate \cite{song2015sinusoidal}.

%%%%%%%%%%%%%%%%%%%%%%%%%%%%%%%%%%%%%%%%%%%%%%%%%%%%%%%%%%%%%%%%%%%%%%%%%%%%%%%%%%%%%%%%%%%%%%%%%%%%

%%%%%%%%%%%%%%%%%%%%%%%%%%%%%%%%%%%%%%%%%%%%%%%%%%%%%%%%%%%%%%%%%%%%%%%%%%%%%%%%%%%%%%%%%%%%%
%\section{Vesicle speed calculated by scaling force with adhered area}
%\begin{figure}[h!]
%\centering
%\includegraphics[scale=0.7]{t_vs_speed_scaled_by_area_sinusoidal_zm_2_ym_30_peak_F_4.eps}
%\includegraphics[scale=0.7]{t_vs_speed_scaled_by_area_sinusoidal_zm_1_ym_15_peak_F_4.eps}
%\caption{Calculation of speed of vesicle by scaling the magnitude of force with adhered area. Parameters are the same as Fig. 3 of the main paper. }
%\label{vesicle_scaled_by_ad}
%\end{figure}

%%%%%%%%%%%%%%%%%%%%%%%%%%%%%%%%%%%%%%%%%%%%%%%%%%%%%%%%%%%%%%%%%%%%%%%%%%%%%%%%%%%%%%%%%%%%%%

%\section{Cell motility on 2D sinusoidal surface}
%\begin{figure}[h!]
%\centering
%\includegraphics[scale=0.7]{trajectory_crescent_along_x_with_background.eps}~~~~~~~~~~
%\includegraphics[scale=0.7]{trajectory_crescent_along_y_with_background.eps}
%\includegraphics[scale=0.7]{trajectory_crescent_along_xy_with_background.eps}
%\caption{Trajectories of vesicle for a 2 dimensional sinusoidal surface. We shoot the motile vesicle in different possible direction, starting on a saddle region.}
%\label{2d-sin}
%\end{figure}

\section*{Supplementary movies}
High resolution movies are also available in thi link {\textcolor{blue}{https://app.box.com/s/fy9sd5ku6msdqa33qbndes02ly9hx1aa}.

\begin{itemize}

\item{\textbf{Movie-S1} (Fig. S-1) Preparation of vesicle on curved substrate. For sinusoidal substrate, we vary  $z_m$ from $1-10~l_{min}$ while keeping $y_m$ fixed  ($=120~l_{min}$). The other parameters are: $N=3127$, $E_{ad}=2.0~k_BT$, $F=2.0~k_BT/l_{min}$,  and $\rho=2.4~\%$.  }
\label{movie1}

\item{\textbf{Movie-S2} (Fig. 2A) Small sized vesicle (N=607) migrating along the axis of sinusoidal substrate starting from the minimum (groove) of the substrate. For sinusoidal substrate, we use $z_m=10~l_{min}; y_m=120~l_{min}$. The other parameters are: $E_{ad}=3.0~k_BT$, $F=4.0~k_BT/l_{min}$,  and $\rho=4.9~\%$.  }
\label{movie2}

\item{\textbf{Movie-S3} (Fig. 2B) Small sized vesicle (N=607) migrating along the axis of sinusoidal substrate starting from the maximum (ridge) of the substrate. For sinusoidal substrate, we use $z_m=10~l_{min}; y_m=120~l_{min}$. The other parameters are: $E_{ad}=3.0~k_BT$, $F=4.0~k_BT/l_{min}$,  and $\rho=4.9~\%$.  }
\label{movie3}

\item{\textbf{Movie-S4} (Fig. 2C) Large sized vesicle (N=3127) migrating along the axis of sinusoidal substrate starting from the minimum of the substrate. For sinusoidal substrate, we use $z_m=10~l_{min}; y_m=120~l_{min}$. The other parameters are: $E_{ad}=2.0~k_BT$, $F=2.0~k_BT/l_{min}$,  and $\rho=2.4~\%$.  }
\label{movie4}

\item{\textbf{Movie-S5} (Fig. 2D) Large sized vesicle (N=3127) migrating along the axis of sinusoidal substrate starting from the maximum of the substrate. For sinusoidal substrate, we use $z_m=10~l_{min}; y_m=120~l_{min}$. The other parameters are: $E_{ad}=2.0~k_BT$, $F=1.0~k_BT/l_{min}$,  and $\rho=2.4~\%$.  }
\label{movie5}

\item{\textbf{Movie-S6} (Fig. S-2A) Large sized vesicle (N=3127) migrating along the axis of sinusoidal substrate starting from the minimum of the substrate. For sinusoidal substrate, we use $z_m=10~l_{min}; y_m=120~l_{min}$. The other parameters are: $E_{ad}=2.0~k_BT$, $F=1.0~k_BT/l_{min}$,  and $\rho=2.4~\%$.  }
\label{movie6}

\item{\textbf{Movie-S7} (Fig. S-2B) Small sized vesicle (N=607) migrating along the axis of sinusoidal substrate starting from the minimum of the substrate. For sinusoidal substrate, we use $z_m=10~l_{min}; y_m=120~l_{min}$. The other parameters are: $E_{ad}=3.0~k_BT$, $F=2.0~k_BT/l_{min}$,  and $\rho=4.9~\%$.  }
\label{movie7}

\item{\textbf{Movie-S8} (Fig. S-2C) Small sized vesicle (N=607) migrating along the axis of sinusoidal substrate starting from the maximum of the substrate. For sinusoidal substrate, we use $z_m=10~l_{min}; y_m=120~l_{min}$. The other parameters are: $E_{ad}=3.0~k_BT$, $F=2.0~k_BT/l_{min}$,  and $\rho=4.9~\%$.  }
\label{movie8}

\item{\textbf{Movie-S9} (Fig. 3A) Small sized vesicle (N=607) migrating orthogonal to the axis of sinusoidal substrate with small wavelength. For sinusoidal substrate, we use $z_m=1~l_{min}; y_m=15~l_{min}$. The other parameters are: $E_{ad}=3.0~k_BT$, $F=4.0~k_BT/l_{min}$,  and $\rho=4.9~\%$.  }
\label{movie9}

\item{\textbf{Movie-S10} (Fig. 3B) Small sized vesicle (N=607) initiating from the maximum of the sinusoidal substrate of small wavelength finally becomes orthogonal to the sinusoidal axis. For sinusoidal substrate, we use $z_m=2~l_{min}; y_m=30~l_{min}$. The other parameters are: $E_{ad}=3.0~k_BT$, $F=4.0~k_BT/l_{min}$,  and $\rho=4.9~\%$.  }
\label{movie10}

\item{\textbf{Movie-S11} (Fig. S-4A) Small sized vesicle (N=607) migrating on sinusoidal substrate at an angle $\sim~45^\circ$ to the sinusoidal axis. For sinusoidal substrate, we use $z_m=1~l_{min}; y_m=15~l_{min}$. The other parameters are: $E_{ad}=3.0~k_BT$, $F=4.0~k_BT/l_{min}$,  and $\rho=4.9~\%$. }
\label{movie11}

\item{\textbf{Movie-S12} (Fig. S-4B) Small sized vesicle (N=607) migrating on sinusoidal substrate along the sinusoidal axis. For sinusoidal substrate, we use $z_m=1~l_{min}; y_m=15~l_{min}$. The other parameters are: $E_{ad}=3.0~k_BT$, $F=4.0~k_BT/l_{min}$,  and $\rho=4.9~\%$. }
\label{movie12}

\item{\textbf{Movie-S13} (Fig. S-4C) Small sized vesicle (N=607) migrating on sinusoidal substrate initially along the sinusoidal axis finally migrates at an angle $\sim~45^\circ$. For sinusoidal substrate, we use $z_m=2~l_{min}; y_m=30~l_{min}$. The other parameters are: $E_{ad}=3.0~k_BT$, $F=4.0~k_BT/l_{min}$,  and $\rho=4.9~\%$. }
\label{movie13}

\item{\textbf{Movie-S14} (Fig. S-4D) Small sized vesicle (N=607) migrating on sinusoidal substrate initially along the sinusoidal axis finally becomes orthogonal to the sinusoidal axis. For sinusoidal substrate, we use $z_m=2~l_{min}; y_m=30~l_{min}$. The other parameters are: $E_{ad}=3.0~k_BT$, $F=4.0~k_BT/l_{min}$,  and $\rho=4.9~\%$. }
\label{movie14}

\item{\textbf{Movie-S15} (Fig. 3C) Karatocytes migrating on sinusoidal substrate at an angle. }
\label{movie15}

\item{\textbf{Movie-S16} (Fig. 3D) Karatocytes migrating on sinusoidal substrate, initially along the axis changes its direction. }
\label{movie16}

\item{\textbf{Movie-S17} (Fig. S-5A) Karatocytes migrating on sinusoidal substrate along the axis maintains its direction of migration. }
\label{movie17}

\item{\textbf{Movie-S18} (Fig. S-5B) Karatocytes migrating on sinusoidal substrate in the orthogonal direction to the sinusoidal axis. }
\label{movie18}

\item{\textbf{Movie-S19} (Fig. S-5C) Karatocytes migrating on sinusoidal substrate almost orthogonal direction to the sinusoidal axis. }
\label{movie19}

\item{\textbf{Movie-S20} (Fig. S-5D) Karatocytes migrating on sinusoidal substrate closely orthogonal direction to the sinusoidal axis. }
\label{movie20}

\item{\textbf{Movie-S21} (Fig. 4A) Vesicle migrating along the axis of a cylindrical fiber, finally migrates in the circumferential direction. Parameters are: $R=10~l_{min}$, $E_{ad}=1.0~k_BT$, $F=2.0~k_BT/l_{min}$,  and $\rho=2.4~\%$. }
\label{movie21}

\item{\textbf{Movie-S22} (Fig. 4G) Overlay of the first micrograph of a time series of migrating cells and
the corresponding 43 trajectories extracted from this series. We used AX2 cells labeled with LimE-GFP. The diameter of the optical fiber is $160 \mu m$ and the images recorded with an Olympus confocal laser scanning microscope Fluoview 1000 in the DIC mode. The different colours of the trajectories are
used to distinguish between individual trajectories. Gray scale represents the DIC intensity in a.u., while the black scale bar corresponds $100 \mu m$. }
\label{movie22}

\item{\textbf{Movie-S23} (Fig. 4J) $2D$ time-lapse projection of a MDCK-LifeAct-GFP cell on a microfiber of $50 \mu m$ in diameter, showing preferential migration along the circumferential axis of the fiber. L-axis and C-axis indicate the cylindrical longitudinal axis and circumferential axis, respectively. Magenta line indicates the cell trajectory. }
\label{movie23}

\item{\textbf{Movie-S24} (Fig. 5A, top panel) D.d.. cells migrating on micropillars of circular cross-section. }
\label{movie24}

\item{\textbf{Movie-S25} (Fig. 5A, bottom panel) D.d.. cells migrating on micropillars of triangular cross-section. }
\label{movie25}

\item{\textbf{Movie-S26} (Fig. 5F) Vesicle migrating on an elliptical cylinder. Parameters are: $R_x=12~l_{min}$, $R_y=7.773~l_{min}$, $E_{ad}=1.5~k_BT$, $F=2.0~k_BT/l_{min}$,  and $\rho=2.4~\%$. }
\label{movie26}

\item{\textbf{Movie-S27} (Fig. S-9A) Vesicle migrating on an elliptical cylinder. Parameters are: $R_x=11.5~l_{min}$, $R_y=8.377~l_{min}$, $E_{ad}=1.5~k_BT$, $F=2.0~k_BT/l_{min}$,  and $\rho=2.4~\%$. }
\label{movie27}

\item{\textbf{Movie-S28} (Fig. S-9C) Vesicle migrating on an elliptical cylinder. Parameters are: $R_x=14~l_{min}$, $R_y=4.882~l_{min}$, $E_{ad}=1.5~k_BT$, $F=2.0~k_BT/l_{min}$,  and $\rho=2.4~\%$. }
\label{movie28}

\item{\textbf{Movie-S29} (Fig. 6A) Vesicle migrating inside a cylindrical tube. Parameters are: $R=35~l_{min}$, $E_{ad}=1.0~k_BT$, $F=2.0~k_BT/l_{min}$,  and $\rho=2.4~\%$. }
\label{movie29}

\item{\textbf{Movie-S30} (Fig. S-11) Non-migrating vesicle spreading and elongating inside a cylindrical tube. Parameters are: $R=23~l_{min}$, $E_{ad}=1.0~k_BT$, $F=2.0~k_BT/l_{min}$,  and $\rho=2.4~\%$. }
\label{30}

\item{\textbf{Movie-S31} (Fig. 6G) $2D$ time-lapse projection of a MDCK-LifeAct-GFP cell in a microtube of $65~\mu m$ in diameter, showing preferential migration along the longitudinal axis of the tube. L-axis and C-axis indicate the cylindrical longitudinal axis and circumferential axis, respectively. Magenta line indicates the cell trajectory. }
\label{movie31}

\end{itemize}

\end{document}